\documentclass[11pt,a4paper]{article}
\usepackage[T1]{fontenc}
\usepackage[utf8]{inputenc}
\usepackage{lmodern}
\usepackage[margin=23mm]{geometry}
\usepackage{amsmath,amssymb,graphicx,booktabs,longtable,array,tabularx}
\usepackage{caption,microtype,xcolor,placeins,listings,etoolbox}

  \usepackage[mathlines]{lineno}
  
\makeatletter
\@ifpackagelater{longtable}{2025/12/22}{}{%
  \patchcmd{\LT@output}{\vss}{\vskip\z@\@plus1fil\@minus\normalbaselineskip}
    {}{\PackageError{longtable-compat}{First output branch not found}{Check the installed longtable version.}}%
  \patchcmd{\LT@output}{\vss}{\vskip\z@\@plus1fil\@minus\normalbaselineskip}
    {}{\PackageError{longtable-compat}{Second output branch not found}{Check the installed longtable version.}}%
}
\makeatother

\usepackage[colorlinks=true,linkcolor=blue!45!black,citecolor=blue!45!black,urlcolor=blue!45!black]{hyperref}
\newcolumntype{Y}{>{\raggedright\arraybackslash}X}
\newcolumntype{R}[1]{>{\raggedright\arraybackslash}p{#1}}
\newcommand{\FeeBL}{\textsc{FeeBL}}
\newcommand{\GP}{\textsc{Guinea-Pig}}
\newcommand{\CAIN}{\textsc{Cain}}
\newcommand{\GeV}{\,\mathrm{GeV}}
\newcommand{\MeV}{\,\mathrm{MeV}}
\newcommand{\ee}{\,e^{+}e^{-}}
\newcommand{\vca}{\theta_{c}/2}

\newcommand{\urad}{\,\mu\mathrm{rad}}
\newcommand{\code}[1]{\texttt{\detokenize{#1}}}
\title{Beam-beam effects on measurements at electron-positron colliders: LEP neutrino counting and Higgs factories}

\author{%
  Brendon Madison\,%
  \href{https://orcid.org/0000-0002-3438-4845}{%
    \includegraphics[height=1.8ex]{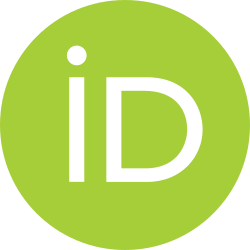}}\\[5pt]
  {\small Deutsches Elektronen-Synchrotron DESY}\\
  {\small Notkestr.~85, 22607 Hamburg, Germany}\\[3pt]
  {\small\href{mailto:brendon.madison@desy.de}
    {\texttt{brendon.madison@desy.de}}}%
}

\date{\today}
\begin{document}
\maketitle

\begin{abstract}
Beam-beam electromagnetic interactions affect the luminosity, collision energy, polarization, and particle backgrounds relevant to precision measurements at electron-positron colliders. We investigate these effects using Fast $e^{\pm}e^{\pm}$ Beam and Luminosity (\FeeBL\ ), a new beam-beam simulation framework, and compare its predictions with published results from LEP and proposed future colliders. We reproduce the analytical field-integration calculations of LEP Bhabha scattering and identify an approximately 2\% bias in previously published beam-deflection simulations. Applying the corresponding luminosity correction and a revised uncertainty assessment motivated by the observed agreement among the four LEP experiments yields ($N_\nu = 2.9965 \pm 0.0040$), compared with the previously reported ($2.9963 \pm 0.0074$). The revised uncertainty is conditional on an assumed common rescaling of experiment specific uncertainties and motivates further investigation of the original uncertainty budgets. We identify and correct implementation errors in the modeling of Breit-Wheeler, Bethe-Heitler, and Landau-Lifshitz pair production, with implications for detector backgrounds at Higgs factories. Simulated beam depolarization and luminosity spectra are compared with existing ILC and HALHF predictions. Finally, we investigate collision energy monochromatization for direct Higgs production, obtaining sub-MeV collision energy peak widths in exploratory linear collider configurations at substantial luminosity cost, while finding significantly weaker narrowing for the FCC-ee configurations studied. We derive a rigid Gaussian beam overlap model and estimate that FCC-ee needs roughly 10 m of horizontal only dispersion to reach widths of $\approx1~\mathrm{MeV}$.
\end{abstract}
\clearpage
\begingroup\small\setcounter{tocdepth}{2}\tableofcontents\endgroup
\clearpage
\section{Scope of simulation}
In $\ee$ colliders, intense electron and positron bunches and fields focus and disrupt each other. The fields cause beamstrahlung, secondary-particle production,
spin precession, and spin-flip transitions. The fields also deflect out-going particles, be they from the bunches or physics processes like Bhabha scattering, at levels that may influence observables. For example, this altered the small-angle Bhabha rate used for luminosity determination at LEP~\cite{lep}. 
At a collider with polarized beams, the luminosity-weighted polarization
also affects polarized cross sections. The beam energy spread of the beams and the collision energy can affect threshold or resonance scans. Beam energy spread and beamstrahlung also enter precision determinations of the collision energy using dimuon events~\cite{madison2022}. For reasons like these, there is a need at $\ee$ collider experiments for fast and precise beam and beam-beam simulations.

Fast $e^{\pm}e^{\pm}$ Beam and Luminosity (\FeeBL) extends upon the
\GP++ code lineage and retains
the original authors' attribution~\cite{guineapig,guineapigpp}. Beam-beam simulations are implemented by representing each bunch by weighted macroparticles in longitudinal field slices. Successive slice encounters use transverse meshes and a fast Fourier transform field solver; additional transverse meshes extend field evaluation beyond the central mesh. The simulations in this paper used both a multi-threaded CPU implementation as well as a single GPU implementation, fixed meshes, and no adaptive integration or kernels. Adaptive meshes, error-aware adaptive families of integration methods and kernels, as well as multiple GPU implementations are currently in development. Fast simulation to interface with event generators, using the methods of GP2X but re-implemented and improved for \FeeBL\!~, are also in development~\cite{gp2x}. 
Simulation settings, such as radiation and secondary processes, are enabled individually by an input card. The physical approximations include locally constant fields approximation (LCFA) for the radiation
kernel and, by default, collinear photon emission. A complete accelerator lattice, ring equilibrium, and detector response are outside this collision model and must currently be handled by other programs.

The comparisons use published \GP, \GP++, \CAIN\ values, and only in the case of \GP++ have new runs been completed. Where original particle records are not supplied, we reconstruct Gaussian bunches from published moments. 
This permits useful consistency and software tests and comparisons, but does not guarantee that non-linear and non-Gaussian properties, like the low energy tails of luminosity spectra, angular correlations, or imported beam structures, can be tested and compared with high precision.
Therefore, we warrant this study as such and leave agreement and numerical convergence as separate studies.

The study of monochromatization is restricted to horizontal dispersion.
Vertical dispersion, longitudinal correlations, and chromatic waist matching
are separate future studies, not mechanisms validated by the results here.
For transparency, the colliders studied here are the International Linear Collider (ILC), the
Linear Collider Facility (LCF), the Hybrid Asymmetric Linear Higgs Factory
(HALHF), and the electron-positron Future Circular Collider (FCC-ee).

\begin{table}[htbp]
\centering\small
\caption{Benchmarks and the subject that each attempts to address.}
\begin{tabularx}{\textwidth}{@{}R{.18\textwidth}R{.21\textwidth}Y@{}}
\toprule
Family & Setting & Principal observables\\
\midrule
LEP & Z-pole bunches & Bhabha deflection, acceptance bias\\
ILC & $500\GeV$ & Incoherent pair yields and luminosity weighted polarization and spin tracking\\
ILC & $1\,\mathrm{TeV}$ & Full luminosity spectra modeling\\
HALHF & $250$, $380$, $550\GeV$ & Full luminosity spectra and beam pair modeling\\
Horizontal dispersion & $125\GeV$ & Collision energy peak width and instantaneous luminosity changes for ILC/LCF, HALHF and FCC-ee\\
\bottomrule
\end{tabularx}
\end{table}

\subsection{Notation, coordinates, units and observables}

We use invariant mass as $W$ as well as collision energy, $W_0$, and center-of-mass energy, $\sqrt{s}$, in this document, and an attempt has been made to ensure that they are not used interchangeably within one context to minimize confusion. For uncertainty assessment, we commonly use the standard error of the mean (SEM) and often refer to it simply as SEM. Similarly, for measurements of the same quantity across different seeds of a simulation, we refer to this as seed SEM. We use natural units in kinematic equations but retain explicit units in inputs and
tables. The longitudinal axis is $+z$; the two incoming beams
travel along $+z$ and $-z$. Beam 1 and beam 2 are identified
explicitly in each input card. A `local angle' about a beam direction is distinct
from an angle about the common $+z$ axis and is noted when used. For beams with crossing angles, the full crossing angle is
$\theta_c$, so equal axis beam energies of $E_0$ give
$W_0=2E_0\cos(\theta_c/2)$ in the ultrarelativistic limit. For unequal beams, like HALHF, with beam energies of $E_{01}$ and $E_{02}$ we find $W_0=2\sqrt{E_{01}E_{02}}\cos(\theta_c/2)$.

By the normalized emittance convention, and with zero dispersion, the transverse size at a waist follows as
\begin{equation}
 \sigma_{x,y}^2=\epsilon_{n,x,y}\beta_{x,y}/\gamma.
\end{equation}
Published emittance and beta function values from references are retained when rounded tabulated sizes differ from this relation. 
The bunch population is a number of physical particles; the macroparticle count is separate and only determines sampling. 
Their ratio sets a macroparticle weight, but generally we will not mention this distinction so it is noted here. Similarly, counts and yields are not true physical counts and yields and must be scaled by their weights. For example, when comparisons of different runs of yields are compared later, they are normalized by their weights to ensure comparisons are unweighted and residuals are not sensitive to differences in weights and are only sensitive to what is being investigated.

For positive collision weights $w_k$,
\begin{equation}
 \mathcal L_{\rm BX}=\sum_k w_k,\qquad
 \mathcal L=f_{\rm BX}\mathcal L_{\rm BX}.
 \label{eq:luminosity}
\end{equation}
When the bunch crossing luminosity, $\mathcal L_{\rm BX}$, is in $\mathrm{m}^{-2}$, the reported instantaneous luminosity
in $\mathrm{cm}^{-2}\mathrm{s}^{-1}$ is
$10^{-4}f_{\rm BX}\mathcal L_{\rm BX}$. Here
$f_{\rm BX}=f_{\rm rep}n_b$ counts bunch crossings per second, which may not be equivalent to accelerator crossings under some circumstances. Beam pair counts are per crossing and per weight; reference production comparisons divide them by the matching electron-positron luminosity to obtain effective daughter-production cross sections. 
The high-energy, peak region, selections of luminosity spectra used here are
$W\geq0.98W_0$ or $W\geq0.99W_0$, with no upper bound. These selections are energy
windows, not the highest two or one percent of events ranked by energy.

The peak response, complete-spectrum RMS, and observed full width at half
maximum (FWHM) are also distinct metrics used for luminosity spectra. Radiation tails can dominate the luminosity spectra RMS even
when the peak is narrow. We use the beta convolved Gaussian response model in
Sec.~\ref{sec:mono} when its extraction passes the stated numerical checks,
and label a directly measured half-height width as an observed FWHM.

For spin expectation $\boldsymbol s$ and particle direction
$\widehat{\boldsymbol p}$, the local helicity is
$h=\boldsymbol s\cdot\widehat{\boldsymbol p}$. Positive helicity denotes something that is right hand dominant, $P=(N_R-N_L)/(N_R+N_L)$. For an accepted collision sample $A$, we use expectation notation and write
\begin{equation}
 \qquad P_{-,A}=\langle h_-\rangle_A,
 \quad P_{+,A}=\langle h_+\rangle_A.
 \label{eq:pol}
\end{equation}
We note that these expectation values are weighted, which means, in this context, luminosity weighted. We distinguish this from an experimental setting where cross section ratios can be used like in Eqs.~(2)-(5) of Ref.~\cite{bmpol}. The experimental measurements are often both luminosity weighted, since the measured cross sections are luminosity weighted, and acceptance weighted given the acceptance of the cross section measurement. We quote polarization changes as percentages of unit polarization: an absolute change of 0.001 is written as 0.1\%. A relative loss instead divides the change by the initial polarization and is identified explicitly. The spin benchmarks start with unit spin along each beam's nominal propagation axis.

For comparison with the explicitly defined 2009 result~\cite{spin2008}, let
$q_1=s_{z,1}$ and $q_2=-s_{z,2}$, and define
\begin{equation}
 \Delta P_{z,L}=100\left[1-\sqrt{\langle q_1q_2\rangle_L}\right]\%.
 \label{eq:axispol}
\end{equation}
Here the numerical value is expressed in percent and the required normalization of 100.
Replacing $q_i$ by $h_i$ defines the local-helicity product loss.
Neither is the conventional effective polarization of a left-right
asymmetry. For annihilation with only the LR and RL initial channels,
the luminosity weighted coefficient is~\cite{polarizedbeams}
\begin{equation}
 P_{{\rm eff},A}=
 \frac{\langle h_-\rangle_A-\langle h_+\rangle_A}
 {1-\langle h_-h_+\rangle_A}.
 \label{eq:peff}
\end{equation}
It is a ratio of averaged cross section coefficients. In the ideal
fully polarized limit, $P_-=\pm1,P_+=\mp1$ and the LR/RL
effective polarization is $\pm100\%$. Outgoing bunch and downstream polarimeter averages require their own definitions and studies done elsewhere~\cite{beckmann}. 

\subsection{Beam pair production and corrections to \GP++}
\label{sec:pair-models}
Beam-induced electrons and positrons, often referred to as beam pairs, contribute to detector occupancy over a wide range of transverse momentum and polar angle. 
Their prediction requires both the cross section and the differential cross sections to be
correct before transport through the bunch fields. Here we focus on the
Breit-Wheeler channel, $\gamma\gamma\to e^+e^-$, the Bethe-Heitler channel,
$\gamma e^\pm\to e^\pm e^+e^-$, and the fully electronic Landau-Lifshitz channel,
$e^+e^-\to e^+e^-e^+e^-$. These follow the beam-background convention
of Ref.~\cite{pairs}.

We inspected the inherited \GP++ v1.2.2 implementation and compared individual
formulae and samplers with independent QED calculations. Table~\ref{tab:pair-fixes}
summarizes verified implementation defects and their changes. 
This attribution is specific to the source build of \GP++ v1.2.2 and does not establish which revisions or options were used for every published collider prediction that comes from any build of \GP~or \GP++. For this reason, we cannot conclusively submit an erratum for existing studies, but past studies are encouraged to redo their beam pair calculations with \FeeBL.

\begin{table}[htbp]
\centering\small
\caption{Verified implementation defects and their changes.
Angular sampling, dispatch and spectator counting changes
apply to pair modes 1 and 2. The normalization correction applies to the
Breit-Wheeler kernel; exact Bethe-Heitler and Landau-Lifshitz
amplitudes carry their own normalization. In mode 2 the inexpensive
Breit-Wheeler estimate controls sampling, with a compensating weight.
The listed numerical effects are not predictions for the net change in a collider background. Those must be done separately and per collider.}
\label{tab:pair-fixes}
\begin{tabularx}{\textwidth}{@{}R{.29\textwidth}Y Y@{}}
\toprule
Component & Inherited behaviour & Corrected behaviour\\
\midrule
Born-level normalization & The numerical approximation is 1.319\% too low. & Corrected using the classical electron radius and $\pi$ values.\\[4pt]
Forward angle sampling & Inverse-CDF sampling tolerance truncates the
small-angle distribution. & Sampling changed to logarithmic angular variable with rejection and stable endpoint.\\[4pt]
Landau-Lifshitz switches & Virtual-virtual collisions omitted even when virtual photons have been generated. & Switch fixed but the input must enable virtual photons.\\[4pt]
Scattered beam particles & With \code{beam_pair=1}, the beam pair counting includes
scattered beam particles. Expected Bethe-Heitler and Landau-Lifshitz counts are then artificially large. & Separate scattered beam particles from the daughter particles.\\
\bottomrule
\end{tabularx}
\end{table}

The discovery of these beam pair issues started with the discovery of the forward angle sampling problem. During tests of beam pairs for HALHF, a review of particle records found events that violated energy-momentum conservation as well as events with negative squared mass. Performing a trace of these specific events, and what parts of the \GP++ and thus \FeeBL~ code they were using, exposed the potential problematic code. The derivations and algorithms in these sections were then reviewed with hand calculations as well as symbolic calculations that use Mathematica~\cite{mathematica},
FeynCalc 10.2.1~\cite{feyncalc10,feyncalc102} and
FeynHelpers~\cite{feynhelpers}, with diagrams generated by
FeynArts 3.12~\cite{feynarts}.

\FeeBL\ uses three pair models. Mode 0 retains the EPA production
model from \GP++ but includes the normalization and threshold corrections. Mode 1 adds
the stable forward-angle sampler and the finite-mass term of the
transverse equivalent-photon flux~\cite{budnev1975},
\begin{equation}
 dn_T=\frac{\alpha}{\pi}\frac{dx}{x}\frac{dQ^2}{Q^2}
 \left[\frac{1+(1-x)^2}{2}-(1-x)\frac{Q_{\min}^2}{Q^2}\right],
 \qquad Q_{\min}^2=\frac{m_e^2x^2}{1-x}.
 \label{eq:pair-epa}
\end{equation}
Here $x$ is the photon energy fraction and $Q^2=-q^2>0$.
Omitting the second term is a leading-logarithmic approximation, distinct
from the normalization error and population definitions in
Table~\ref{tab:pair-fixes}.

Mode 2 evaluates the complete lowest-order QED matrix elements for the
Bethe-Heitler and Landau-Lifshitz final states. Exact internal four-momenta and phase-space weights replace the
on-shell $\gamma\gamma$ approximation and its hard \code{pair_q2} veto.
The calculation uses importance sampling. The current host-code mapping to the beam directions still uses a common transverse boost; exact internal kinematics do not remove that
angular approximation. These are tree-level calculations, and should not be treated as high precision and radiatively corrected predictions.

\subsection{Beam pair notation}
Beam pair yields count individual electrons and positrons. We use notation that refers to a ``near axis'' as it ensures that the calculated and presented angles for the beam pairs are clearer. If the common interaction point is used, instead of the near axis, the resulting angles for beam pairs will contain zeroes and divergences in the plotted structures. We define the near axis form for a particle with momentum $\boldsymbol p$,
\begin{equation}
 p_T=\sqrt{p_x^2+p_y^2},\qquad
 \theta_{\rm near}=\operatorname{atan2}(p_T,|p_z|).
 \label{eq:angle}
\end{equation}
The nearest axis angle folds both beam directions into $[0,\pi/2]$ and ensures the $\log_{10}(\theta)$ plots do not form a narrow and vertical feature near $\log_{10}\pi$.

\subsection{Monochromatization notation}
\label{ssec:mono}
For horizontal dispersion, the canonical map is
\begin{equation}
 x\longmapsto x+D\delta,\qquad
 \zeta\longmapsto\zeta+Dp_{x,can.},
 \label{eq:dispersionmap}
\end{equation}
where $p_{x,can.}$ is normalized canonical transverse momentum and
$\delta=(E-E_0)/E_0$. The longitudinal term is required for symplecticity.
In \FeeBL~ dispersion precedes the crab transformation so the opposite case is not currently supported. After these maps, \code{incoming_map_reslice=1} reassigns particles using
their transformed longitudinal coordinates while preserving identity, spin
and weight. A transformed particle outside the declared longitudinal domain
causes setup to fail instead of being silently discarded or clipped.

Imported primary beams can use the native ASCII loading interface as is done in \GP++. A card
marked \code{loaded_beam_is_pretransformed=1} declares that the incoming
transformations have already been applied; it skips crossing, roll, offsets
and incoming maps.

\FloatBarrier

\section{Installation, input cards and output}
\label{sec:manual}
The accompanying repository contains the simulator, analysis tools, and example
cards.

\subsection{Build instructions}
The supported minimal build uses Linux or Windows Subsystem for
Linux (WSL), CMake 3.16 or newer, a C++17 compiler and FFTW3. Python is used for
benchmarks and analysis. The runner requires Python 3.10 or
newer; replotting requires NumPy 1.24 or newer and Matplotlib 3.7 or newer.
SciPy is required for viewer finalization as well as numerical fitting.
The browser viewer is a static
HTML application.
Optional CUDA and parallel backends have separate requirements and are not
needed to reproduce the serial results discussed here.

From the repository root, a standard CPU build is
\begin{lstlisting}
sudo apt-get install build-essential cmake libfftw3-dev \
  libhdf5-dev python3 python3-venv nodejs
python3 -m venv .venv
. .venv/bin/activate
python -m pip install -r requirements-analysis.txt
cmake -S . -B build -G "Unix Makefiles" \
  -DCMAKE_BUILD_TYPE=Release -DFEEBL_ENABLE_CUDA=OFF
cmake --build build --parallel 4
ctest --test-dir build -LE strict --output-on-failure
\end{lstlisting}
The executable is \code{build/src/feebl}. For an FFTW installation outside
the system search path, provide its installation prefix with
\code{-DCMAKE_PREFIX_PATH=/path/to/fftw}. The CMake configuration output
identifies the selected library. Keep that output and the compiler version
with a new benchmark. The Makefiles generator is specified because the
existing spin-test link procedure uses its generated link information.

CPU worker pools are included in the normal build and are selected by the
input card, not by the number of compilation jobs. OpenMP is a separate
build option, \code{-DFEEBL_ENABLE_OPENMP=ON}. CUDA support requires
\code{-DFEEBL_ENABLE_CUDA=ON}, a CUDA toolkit and a supported GPU.
The resident GPU modes impose different process and mesh restrictions. The collider results here use the serial backend unless noted otherwise.

The packaged runner lists the available cases and creates a separate output
directory for a first collision:
\begin{lstlisting}
python benchmarks/run.py list
python benchmarks/run.py run smoke \
  --feebl build/src/feebl --output runs/smoke
python benchmarks/run.py verify
\end{lstlisting}
This test also checks the installation and the output path.

\subsection{Input cards}
Like in \GP++, the input card selects an accelerator block and a parameter block from
\code{acc.dat} in the working directory:
\begin{lstlisting}
feebl --make_gui_files <accelerator_block> <parameter_block> <output_file>
\end{lstlisting}
The accelerator block defines the bunches and optics; the parameter block
selects numerical resolution, processes, and outputs. For complete runnable
cards, use \code{benchmarks/cases/}. The following illustrates the
organization and the opposite dispersion signs; it is an example and not a complete card:

\noindent\begin{minipage}{\linewidth}
\begin{lstlisting}
$ACCELERATOR::example
{
  energy=62.501531;
  particles=2.0;
  beta_x=3.25;
  beta_y=0.41;
  sigma_z=300.0;
}
$PARAMETERS::example_settings
{
  ip_dispersion.1=0.0003712;
  ip_dispersion.2=-0.0003712;
  incoming_map_reslice=1;
}
\end{lstlisting}
\end{minipage}
Native cards retain historical units. In this example, energy is in GeV,
population is in units of $10^{10}$, beta functions are in mm, bunch length
is in $\mu$m and dispersion is in meters. A suffix \code{.1} or \code{.2} sets the
corresponding beam.

\begin{table}[htbp]
\centering\small
\caption{Input groups relevant to the examples. Complete cards determine all
process flags and retain the exact native keyword spellings and units.}
\begin{tabularx}{\textwidth}{@{}R{.27\textwidth}Y@{}}
\toprule
Group & Interpretation and common error to avoid\\
\midrule
Energy, population, emittance and beta & Physical incoming bunch parameters.
Macroparticle count changes sampling, not the requested physical population.\\[3pt]
Energy spread & Applied to primary particles before incoming maps. Enabling
dispersion for a monoenergetic beam does not create the intended energy--position correlation.\\[3pt]
Crossing, crab and waist settings & Geometric crossing, RF-like bunch tilt and
crab-waist optics are distinct. Match signs and map order for both beams.\\[3pt]
\code{ip_dispersion.1}, \code{ip_dispersion.2} & Horizontal IP dispersions in m.
The canonical map changes both $x$ and the longitudinal coordinate.\\[3pt]
\code{incoming_map_reslice} & Reassigns slices after canonical incoming maps.
Required by the high-dispersion examples to retain scheduling consistency.\\[3pt]
Particle, mesh, slice and time settings & Numerical controls. Increase each
separately at fixed physical inputs before a combined refinement.\\[3pt]
Spectrum limits and bins & Choose support and peak resolution before running.
A narrow peak cannot be recovered from a single broad energy bin.\\[3pt]
Radiation, spin and secondary switches & Define the physical model. Switching
one process can change subsequent random draws and trajectories.\\
\bottomrule
\end{tabularx}
\end{table}

\paragraph{Incoherent-pair settings.}
The models in Sec.~\ref{sec:pair-models} are selected by
\code{pair_model=0}, 1 or 2. For a pure daughter study, a
parameter block excerpt may look like
\begin{lstlisting}
pair_model=1; do_pairs=1; track_pairs=1;
do_photons=1; do_photon_lumi=1; electron_ratio=1;
beam_pair=0; do_compt=0; do_coherent=0; do_trident=0;
load_events=0; do_bhabhas=0;
store_pairs=2; pair_ratio=1; pair_ecut=0.005;
\end{lstlisting}
The energy cut is in GeV and acts on each daughter at production.
\code{store_pairs=2} preserves birth records in \code{pairs0.dat} and
final records in \code{pairs.dat}; these are two stages of the
same population, not additive samples. A smaller \code{pair_ratio}
reduces storage and requires its inverse weight for per-crossing yields.
Use final admitted records for the comparison without applying the birth
cut a second time. Native production-expectation counters precede admission
and need not equal the stored daughter count. Complete benchmark cards
specify the remaining beam, numerical, and photon settings.

\subsection{\FeeBL\ GUI: beam viewer and diagnostics}
\FeeBL~ has a dedicated beam viewer to help with visualizing results and diagnosing problems. The beam viewer loads a finalized local run directory; an input card alone is
not sufficient to run the beam viewer. For example, below uses the executable and input card, source:
\begin{lstlisting}
python tools/finalize_gui_dataset.py runs/smoke \
  --binary build/src/feebl --card runs/smoke/acc.dat --source .
python tools/serve_viewer.py --run runs/smoke
\end{lstlisting}
The server prints the local browser address and serves only the selected
run and viewer assets. The saved demonstration can be
opened without a new simulation:
\begin{lstlisting}
python tools/serve_viewer.py \
  --run benchmarks/example_output/gui_smoke
\end{lstlisting}
The folder chooser also accepts another finalized directory. The browser
reads those selected files locally. The persistent frame
slider and playback control remain available while scrolling so that the user can see individual step results and total integrated results per their choosing. The beam
direction arrows describe the saved display coordinate convention.

The animation uses a deterministic primary-particle subset for display.
Scientific moments and histograms are read from native accumulators. The selected frame synchronizes the plots. Table~\ref{tab:viewer-scopes} distinguishes their temporal meanings.
\begin{table}[htbp]
\centering\small
\caption{Frame and total modes in the viewer. A frame interval contains
collision contributions since the preceding saved state.}
\label{tab:viewer-scopes}
\begin{tabularx}{\textwidth}{@{}R{.25\textwidth}Y Y@{}}
\toprule
Quantity & Current frame & Total\\
\midrule
Luminosity and radiation spectra & Contributions in the frame interval &
All collision contributions\\
Collision energy and polarization & Luminosity-weighted accepted collisions in
that step & Luminosity-weighted accepted collisions over all steps\\
Beam moments & Unweighted stored primary state & Pooled collision-weighted
first and second moments\\
Secondary kinematics & One stored state per record &
Final stored records\\
Imported particle beam deflection & Change since admission at the selected state &
Change since admission at the final state\\
\bottomrule
\end{tabularx}
\end{table}
The first stored frame already follows a collision interval and is not an
exact pre-collision snapshot. Frame moments use stored slice states, whereas
the displayed primary cloud can drift inactive slices to a common endpoint.
These are different sampling conventions. Total RMS values are formed from
pooled first and second moments, not from averages of frame RMS values.

Nominal spin settings, whole-bunch helicity and luminosity-weighted electron
and positron helicities are shown separately. High-energy means and
polarizations use $W\geq0.98W_0$. The interface does not substitute a
spin-product average for the conventional $P_{\rm eff}$ in
Eq.~\eqref{eq:peff}. A distribution that was not computed is distinguished
from a calculated zero yield. Beamstrahlung and pair-energy histograms have their
own native production weights and are not histograms of the displayed sample.

The longitudinal field overlays use $(z,x)$ at $y=0$ and $(z,y)$ at $x=0$.
They show the transverse electric-equivalent magnitude
$E_\perp=\sqrt{E_x^2+E_y^2}$, converted from native GV/nm to displayed
GV/m. A single beam uses
a Cividis-derived gradient; two partially transparent blue and orange layers
retain the source identity and a common scale. Showing both is not a vector
sum: two scalar magnitudes cannot determine the net field. Fixed physical
map coordinates prevent an apparent contraction from automatic pixel
rescaling. These overlays do not represent a complete laboratory electromagnetic field
or demonstrate external laser-field support.
Figure~\ref{fig:viewer} shows the field controls and persistent playback
in the packaged installation example, followed by frame-resolved numerical
diagnostics from a polarized collision. The latter distinguish the configured
polarizations, current bunch means, and luminosity-weighted collision values.

\begin{figure}[p]
\centering\includegraphics[width=\textwidth]{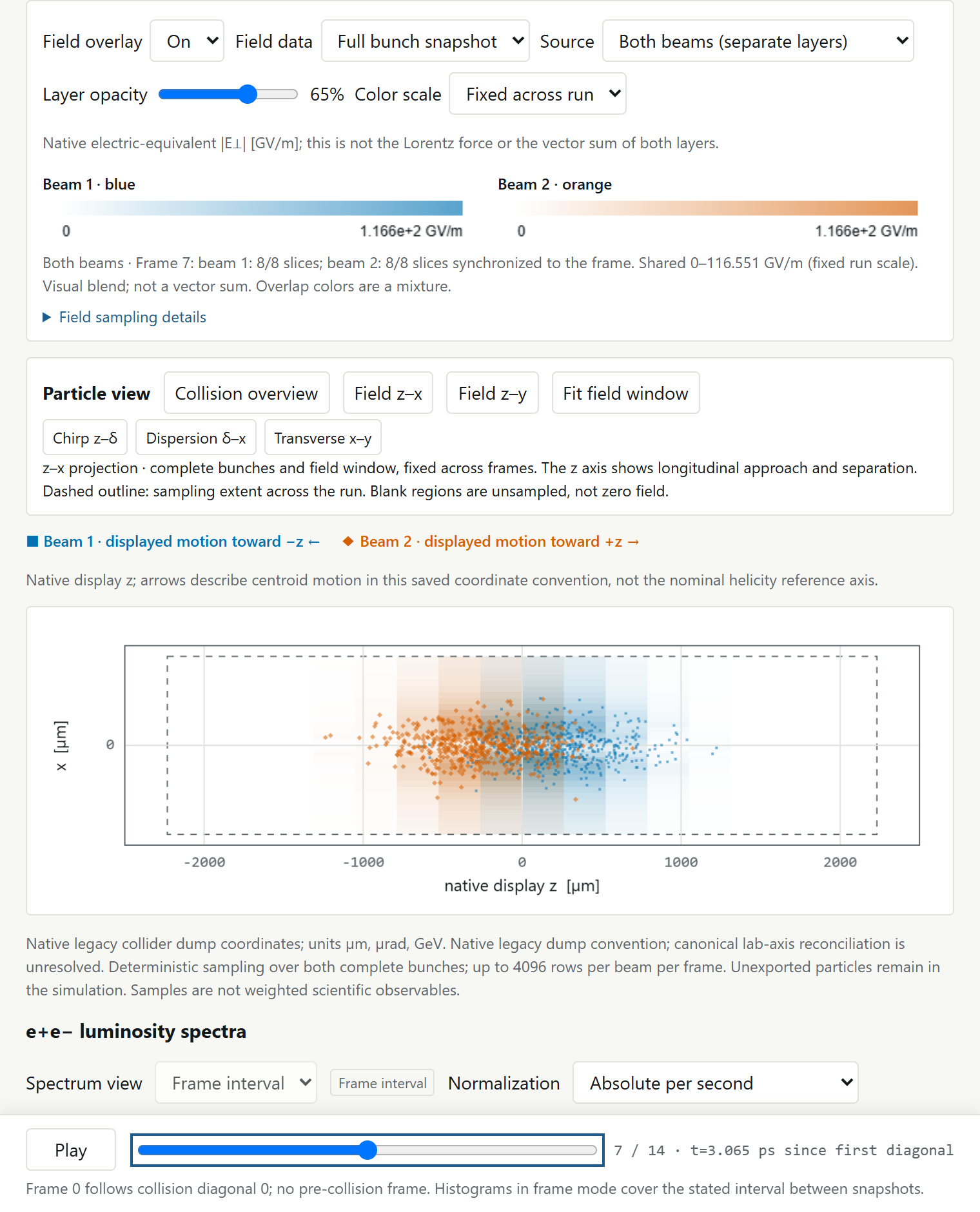}
\caption{\FeeBL\ viewer at saved frame 7 (numbered 0--14) of the 500 GeV
installation example. Blue and orange identify the sampled bunches and
their separate transverse electric-equivalent field layers on a common
GV/m scale. Overlapping colors do not represent a vector sum. The dashed
outline marks the sampling extent. Projection controls and fixed playback
permit frame inspection. This small demonstration is not a converged
collider prediction.}
\label{fig:viewer}
\end{figure}

\begin{figure}[p]
\ContinuedFloat
\centering\includegraphics[width=160mm]{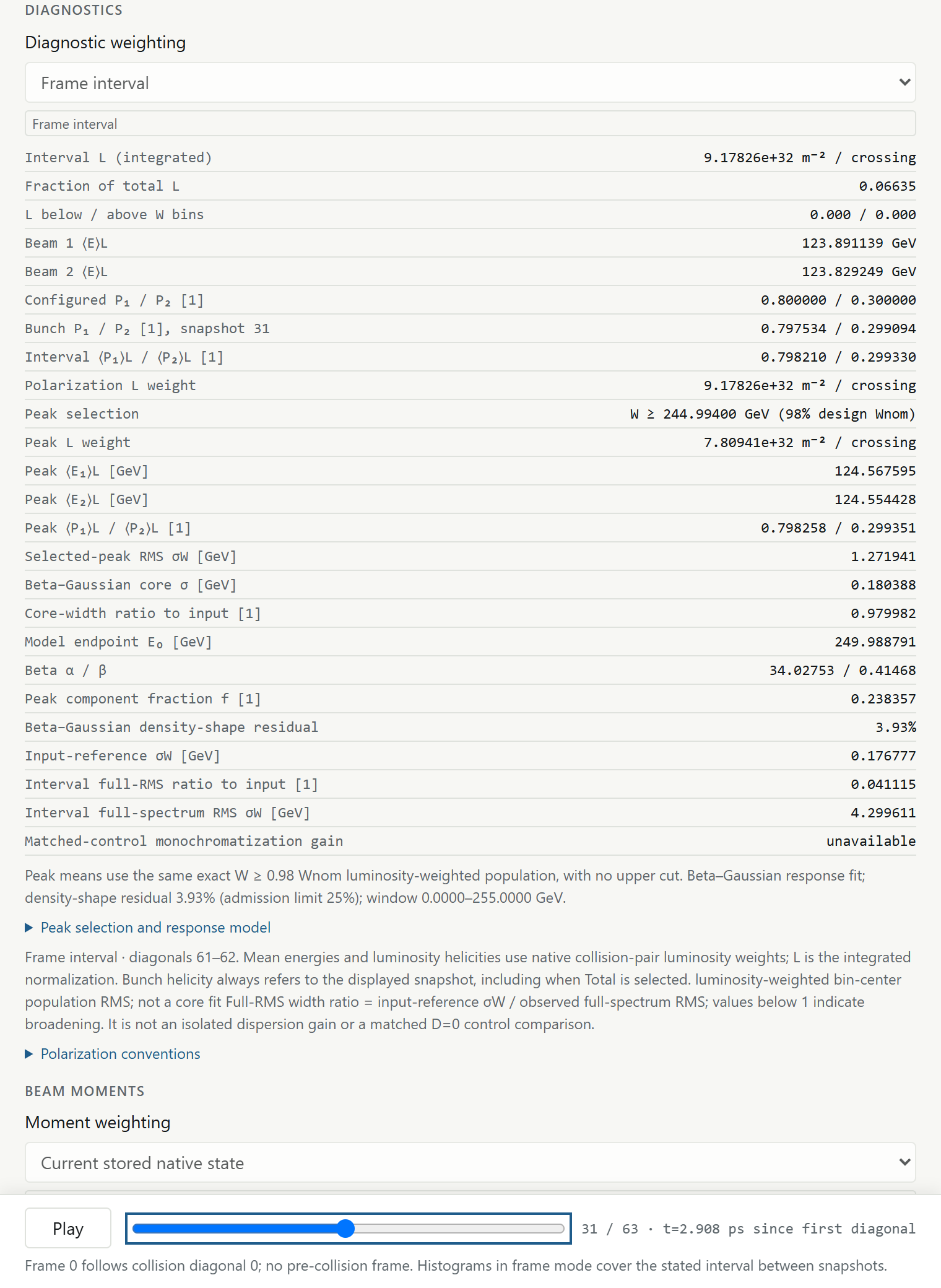}
\caption{Continued: quantitative diagnostics at frame 31 of a saved
250 GeV polarized collision. The frame-interval selector and playback
position remain visible. Energy and polarization means are reported for
the interval and its $W\geq0.98W_0$ subset; the current bunch means use
the displayed snapshot. This preserved example illustrates the interface,
not a precision validation of the displayed fitted response.}
\label{fig:viewer-diagnostics}
\end{figure}

The beam pairs panel plots $\log_{10}(\theta/\mathrm{rad})$ against
$\log_{10}[p_T/(\mathrm{GeV}/c)]$ as is the standard for assessing beam pairs. For a two-body Bhabha beam electromagnetic deflection test, four distributions describe electron and positron
polar and azimuthal changes. They are changes since admission, not changes
between adjacent saved frames.

For supported imported events, \code{gui_imported_trajectories.json} stores
the initial record identity and the direction at admission, each saved frame
and the final state. The finalizer joins persistent particle labels to the
event ID, particle slot and PDG ID in \code{event_identity.json}. It writes
\code{collider_event_deflections.json}, with fixed-bin angular histograms
in degrees, frame timing, admission counts and source hashes. The viewer
downloads both this analysis and the native records. Event matching and
angular extraction therefore require no separate shell sorting script.
This adapter currently supports a pure population of ordinary eight-column
imported events in the serial, head-on tracker, with at most 8192 identified records and 64
saved frames. Its trajectory rows contain direction and field-gather counts;
they do not supply final particle energies or positions.

\FloatBarrier

\section{Implications of beam-beam effects for experimental measurements}
\label{sec:comparisons}
The packaged reference tables contain values extracted from the primary
publications listed in their provenance records. We distinguish direct sampling
errors from estimated curve uncertainties and numerical sensitivity.
The standard error of the mean (SEM) across $R$ independent seeds with a sample standard deviation of $s$ is written here as $s/\sqrt R$. Numerical changes below are controlled sensitivities and are neither confidence intervals nor calibrated total numerical errors.

The production ensembles retain grid and source stencil coverage warnings. Independent-seed errors at a fixed domain do not bound the resulting observable bias. Larger-domain studies at fixed cell spacing remain necessary before precision interpretation. Therefore, we do not claim that high precision runs have been conducted nor that high precision results are achieved. Instead, the level of accuracy consistent with previous beam-beam particle-in-cell simulations is achieved in \FeeBL~ too.

Table~\ref{tab:collider-inputs} gives the principal physical and numerical
inputs for the collider comparisons. Equal beam values are printed once.
The complete evaluated input table and figure to case mapping accompany the
release; these retain energy spreads, emittances, domain sizes, process
switches and reproduction status.
\begin{table}[htbp]
\centering\small
\caption{Selected collider inputs. $N_m$ is macroparticles per beam and
the grid column is $N_x\times N_y\times N_z$. These are simulation settings,
not demonstrated convergence thresholds. HALHF energies are asymmetric so two energies are provided instead of one.}
\label{tab:collider-inputs}
\begin{tabular}{@{}lrrrrrr@{}}
\toprule
Case & $E_{1,2}$ [GeV] & $N_{1,2}/10^{10}$ & $\beta_x$ [mm] & $\beta_y$ [mm] & $N_m$ & Grid\\
\midrule
HALHF 250 spectrum & 375, 41.7 & 1, 3 & 3.3 & 0.1 & 10000 & 256$\times$1024$\times$32\\
HALHF 380 & 570, 63.3 & 1, 3 & 3.3 & 0.1 & 10000 & 128$\times$128$\times$32\\
HALHF 550 & 825, 91.7 & 1, 3 & 3.3 & 0.1 & 10000 & 128$\times$128$\times$32\\
ILC 1 TeV & 500 & 1.74 & 11 & 0.23 & 10000 & 128$\times$128$\times$32\\
LEP focusing & 45.6 & 12.8 & 2000 & 50 & 40000 & 128$\times$256$\times$64\\
ILC pairs & 250 & 2 & 15 & 0.4 & 10000 & 128$\times$128$\times$32\\
HALHF 250 pairs & 375, 41.7 & 1, 3 & 3.3 & 0.1 & 10000 & 128$\times$128$\times$32\\
\bottomrule
\end{tabular}

\end{table}

For graphical spectra and references, digitization was performed on published results and plots, and then simulated weights were transferred to the published
bins by overlap and both distributions are normalized on the common retained
support. Full and peak panels are normalized separately. Thus, the cumulative distribution distance tests their shapes, not probability outside the displayed interval. We acknowledge that there are likely other ways to robustly digitize references, but this is what was used here.

\subsection{Methodology of reference errors and pulls}
For bin contents $s_i,r_i$ from \FeeBL\ and the reference respectively, the displayed pull is
\begin{equation}
 p_i=\frac{s_i-r_i}{\sqrt{\sigma_{s,i}^2+\sigma_{r,i}^2-2\ \mathrm{Cov}(s_{i},r_{i})}}.
 \label{eq:pull}
\end{equation}
Both simulation and reference variance, and their covariance, enter with the covariance evaluated using differences between different parts of runs with the same seed.

Reference-curve fluctuations are estimated before interpolation or
normalization by removing local linear trends~\cite{gasser1986}. Spectra are
first expressed as densities. A three-point local pilot supplies
heteroscedastic scaling; the nearest 15 squared contrast probes in each
contiguous interval are pooled using a median with its Gaussian-noise
calibration. A five-point line fit and 9/23-probe pooling are companion
checks. Curvature, extraction and plotting can contribute to the estimate.
No displayed ordinate is smoothed. Gaps are not bridged, and short support
segments borrow a curve-wide fluctuation scale.

The covariance check reserves eight collisions for training and four
independent collisions for validation. Bin coverage and pull RMS should be treated as descriptive because this digitization and quantization method is not precise nor representative of the original data it approximates.

\subsection{LEP Bhabha beam deflection}
Here we reproduce the LEP Bhabha beam deflection results. We use ten primary bunch realizations, each
with 1024 collision vertices at each initial polar angle, 31.3 and
51.6 mrad. Each event contains an electron and a positron, giving a total of 40960
tracked particles. The initial
transverse boost of Ref.~\cite{lep} is omitted, so these probes approximate
the deflection calculation rather than a complete radiative Bhabha sample.
The identified event interface in Sec.~\ref{sec:manual} exports native
initial and tracked directions as JSON, then produces frame-resolved and
final angular distributions automatically. This replaces the separate
output and then sorting step used in the \GP++ workflow and in
Ref.~\cite[Sec.~4.3.1.1]{madisonthesis}; the new output maintains event identity and coordinate
conventions with the numerical output. This replacement enables better pipelining of data for recent investigations into using event-by-event corrections to beam-deflection biases in
luminosity measurements~\cite{madison2026}.

For deflection of $\Delta\theta=\theta_i-\theta_f$ and an angular rate proportional
to $\theta^{-3}$, the leading acceptance correction is
\begin{equation}
 \frac{\Delta\mathcal L}{\mathcal L}=
 -\frac{2}{\theta_{\min}^{-2}-\theta_{\max}^{-2}}
 \left[\frac{\langle\Delta\theta\rangle_{\min}}{\theta_{\min}^3}
 -\frac{\langle\Delta\theta\rangle_{\max}}{\theta_{\max}^3}\right].
 \label{eq:lepbias}
\end{equation}
Each field supplies the azimuthal means, a conditional focusing histogram
and both boundary means. Their covariance gives the standard error of the
ensemble mean and the uncertainty of Eq.~\eqref{eq:lepbias}. Two matched
zero-field controls check transport separately; they are not subtracted
from unrelated fields. We use the same field integration methods as the published study for an analytic comparison as done in the reference study.

\begin{table}[htbp]
\centering\small
\caption{LEP Bhabha (electron) beam deflection and the difference to the field integration model. \FeeBL\ values are means of ten independent
fields with standard errors. The reference terms are not their full-acceptance results but their simulation results to make the comparison valid. We include their claimed fractional uncertainty as the quoted uncertainty. We observe statistically significant bias in the reference but not in the results for \FeeBL\ .}
\label{tab:lep-focusing-ensemble}
\begin{tabular}{@{}lrrr@{}}
\toprule
Observable & Field Integration & Reference~\cite{lep} & \FeeBL \\
\midrule
Inner deflection [$\mu$rad] & $13.09$ & $12.81\pm0.09$ & $13.12\pm0.08$ \\
Outer deflection [$\mu$rad] & $11.49$ & $11.19\pm0.10$ & $11.49\pm0.06$ \\
Inner Difference [\%] & -- & $-2.2\pm0.5$ & $0.2\pm0.4$\\
Outer Difference [\%] & -- & $-2.6\pm0.6$ & $-0.3\pm0.3$\\
\bottomrule
\end{tabular}

\end{table}
Applying Eq.~\eqref{eq:lepbias} to the published boundary means gives
$-0.10425\%$. By comparison, it is $-0.10671\%$ for the \FeeBL\ results, which is a $\approx2\%$ bias. We cannot determine the origin of this observed $\approx2\%$ deficit using both the analytical model and \FeeBL\ simulation results. For a better comparison, a repository of the original published data or simulation settings is needed. This bias also carries over to the later published value which used improved theory calculations~\cite{theory}.

During the examination of this result, we also examined the published uncertainty assessment. The four LEP determinations of $N_\nu$ agree more closely than their published spreads indicate, leading to a lower-tail probability of $2.05\%$ for the observed scatter in a $\chi^2$ test using the published values and covariances. 
Holding the shared uncertainties fixed and rescaling the experiment specific standard deviations by a common factor places the observed scatter at the
$68.27$th percentile for a combined uncertainty of
$\sigma(N_\nu)=0.00401$. Inverting the same test at the 15.87th and
84.13th percentiles gives the corresponding central $68.27\%$ interval
of $[0.00391,0.00485]$ for the uncertainty. We suspect that over-estimation or mis-estimation in one or more dominant uncertainty sources may be causing this result. 
One dominant source occurs between the hadronic pole cross section, $\sigma^0_{\rm had}$, and the width ratio, $R_\ell=\Gamma_{\rm had}/\Gamma_{\ell\ell}$. This result motivates a smaller uncertainty, near $\sigma(N_\nu)=0.00401$, but it requires that a reassessment of each experiment's uncertainty budget is performed before it can defensibly be claimed.

If we use both the revised $68.27$th percentile uncertainty estimate and the bias correction from \FeeBL\ on the values of the reference's calculation of the number of neutrino species, the results are
\begin{align}
    N_{\nu,ref} = 2.9963~\pm~0.0074\\
    N_{\nu,new} = 2.9965~\pm~0.0040~,
    \label{eq:neu}
\end{align}
a small change relative to the original measurement and its uncertainty.

\begin{figure}[htbp]
\centering\includegraphics[width=\textwidth]{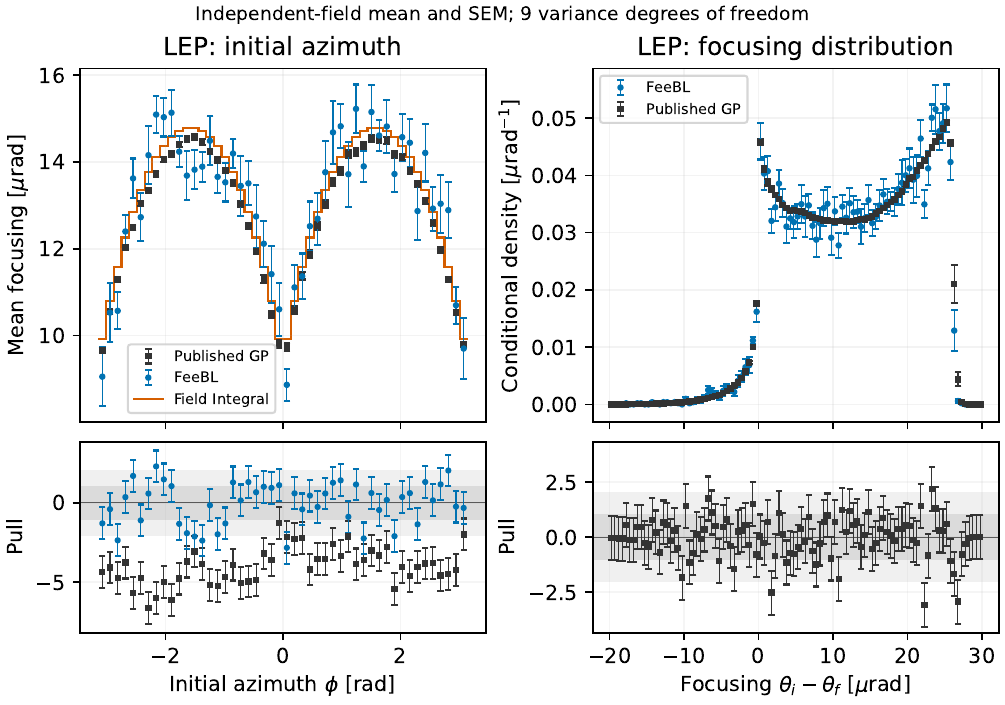}
\caption{LEP Bhabha (electron) beam deflection at an initial polar angle of 31.3 mrad.
Left: mean deflection versus initial azimuth. Right: the density of events per polar angle deflection over a beam crossing with FeeBL result rescaled to match the normalization of the published values. The references are the published \GP\ and
field-integration calculations~\cite{lep}.}
\label{fig:lep}
\end{figure}

In Fig.~\ref{fig:lep}, azimuthal pulls from the published \GP\ values have a mean of -4.00 while \FeeBL\ have a mean of -0.03. These results reinforce the bias found in Table~\ref{tab:lep-focusing-ensemble}, that the published \GP\ values are biased lower than the field-integration result while the same bias is not found in the \FeeBL\ values. We also observe a mean pull for the published \GP\ and \FeeBL\ values for electron polar angle focusing/deflection of -0.01, but only after normalizing both histograms to the same scale. This normalization required scaling the \FeeBL\ results by roughly $-2.49\%$, which is inline with the previous bias in the published \GP\ values found in Table~\ref{tab:lep-focusing-ensemble}. These results support the hypothesis that it is plausible that \FeeBL\ is reproducing the underlying physics correctly, as the shape and analytical comparisons are matched.

\subsection{ILC pair physics and luminosity spectra at 500 GeV and 1 TeV}
The 500 GeV pair configuration uses the parameter set labeled USSC in
Table~I of Ref.~\cite{pairs}: $2\times10^{10}$ particles per bunch,
$\beta_{x,y}=(15,0.4)$ mm and $\sigma_z=300\,\mu$m. We compare the
finite-mass EPA model and the leading order matrix-element model of
Sec.~\ref{sec:pair-models}, with twelve independent collisions per model.
Common finite-beam and photon upgrades make
this a controlled model comparison rather than an exact rerun of the
historical \GP++\ implementation. Only incoherent daughters enter; other
backgrounds and spectator beam leptons are excluded. The 5 MeV selection
acts on each daughter at production.

The EPA calculation uses the published \GP\ prescription
$Q^2_{\max}=s_{\rm pair}/4$ with beam-size suppression, whereas the cited
\CAIN\ calculation uses $Q^2_{\max}=m_e^2$~\cite{pairs}. Mode 2 does not
apply the EPA hard virtuality veto. Following Table II of Ref.~\cite{pairs}, all three
channels are compared as effective daughter production cross sections,
$\sigma_j^{\rm eff}=\sum_a N_{j,a}/\sum_a\mathcal L_{ee,a}$. We suspect that the differences observed between \CAIN\ and \GP\ and \FeeBL\ are from the different treatment of $Q^2_{\max}$ but further tests need to be conducted to confirm this. The following controls separate these effects in the present implementation.

\begin{table}[htbp]
\centering\small
\caption{Effective luminosity weighted daughter production cross sections in mb
for ILC at $500\,\mathrm{GeV}$. Cross section uncertainties are the outer estimated
half-ranges defined in Fig.~\ref{fig:pair-yields}, rather than statistical
standard errors alone.}
\label{tab:ilc-pair-ensemble}
\begin{tabular}{@{}lrrrr@{}}
\toprule
Quantity & \GP & \CAIN & \FeeBL\ model 1 & \FeeBL\ model 2\\
\midrule
Breit--Wheeler & 1.05 & 1.04 & $1.163\pm0.123$ & $1.125\pm0.200$\\
Bethe--Heitler & 37.70 & 34.50 & $38.870\pm1.087$ & $39.187\pm1.100$\\
Landau--Lifshitz & 19.20 & 15.20 & $18.168\pm0.765$ & $18.839\pm0.496$\\
Total & 57.95 & 50.74 & $58.201\pm1.953$ & $59.151\pm1.734$\\
$\mathcal L_{\rm BX}$ [$\mu\mathrm b^{-1}$] & 1.82 & 1.82 & $1.7348\pm0.0116$ & $1.7443\pm0.0095$\\
\bottomrule
\end{tabular}

\end{table}

\begin{figure}[htbp]
\centering\includegraphics[width=.92\textwidth]{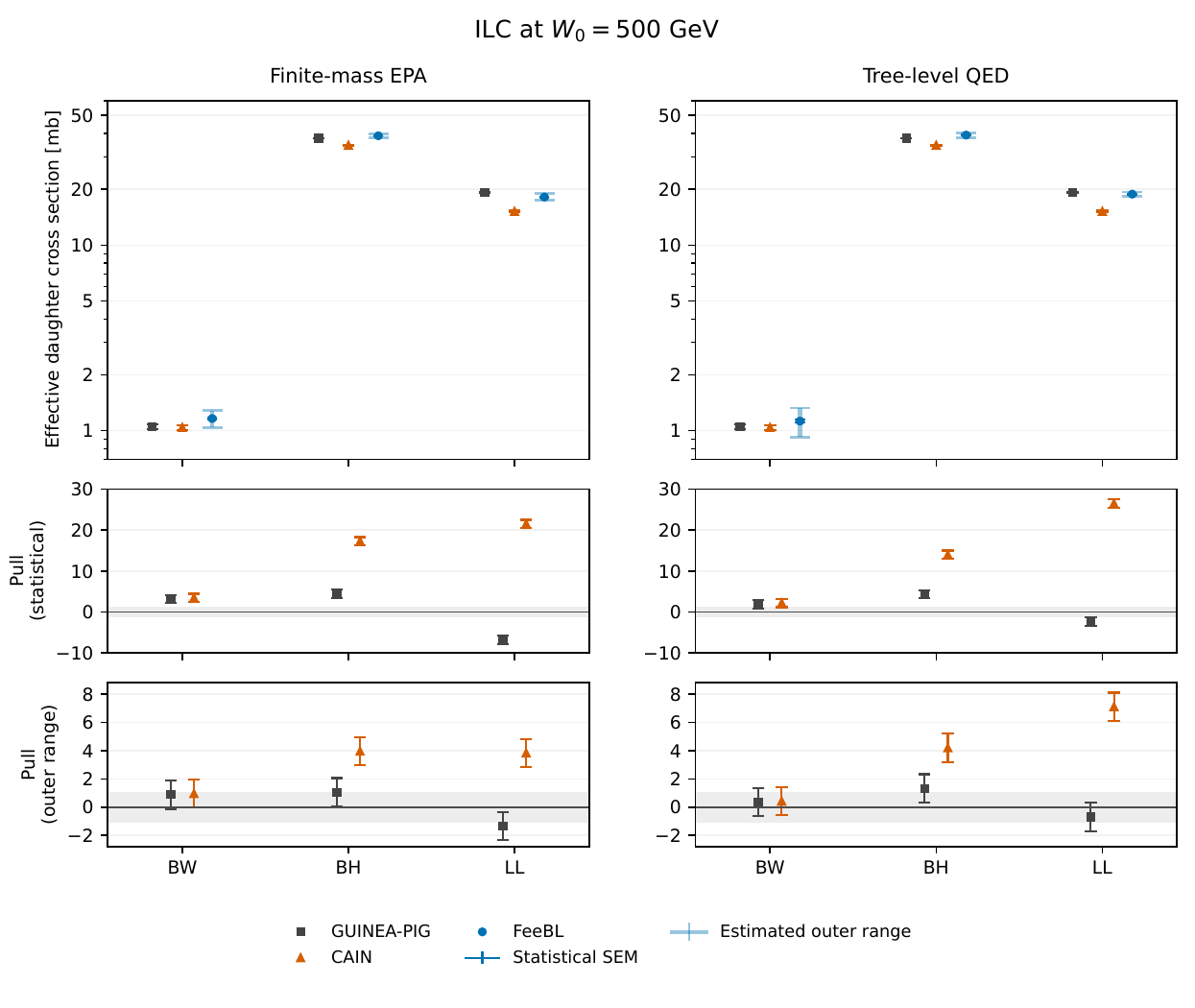}
\caption{ILC effective daughter production cross sections for the
finite-mass EPA model (left) and leading order matrix-element model (right).
Each \FeeBL\ point divides the summed yield by summed electron-positron
luminosity across twelve complete collisions.
The references use different virtual-photon prescriptions~\cite{pairs}.
These scenarios do not recover the original reference simulations
and instead we provide two uncertainty bars and pulls, the first for pure statistical in \FeeBL\ and the latter including other uncertainty sources.}
\label{fig:pair-yields}
\end{figure}

\begin{figure}[p]
\centering\includegraphics[width=\textwidth]{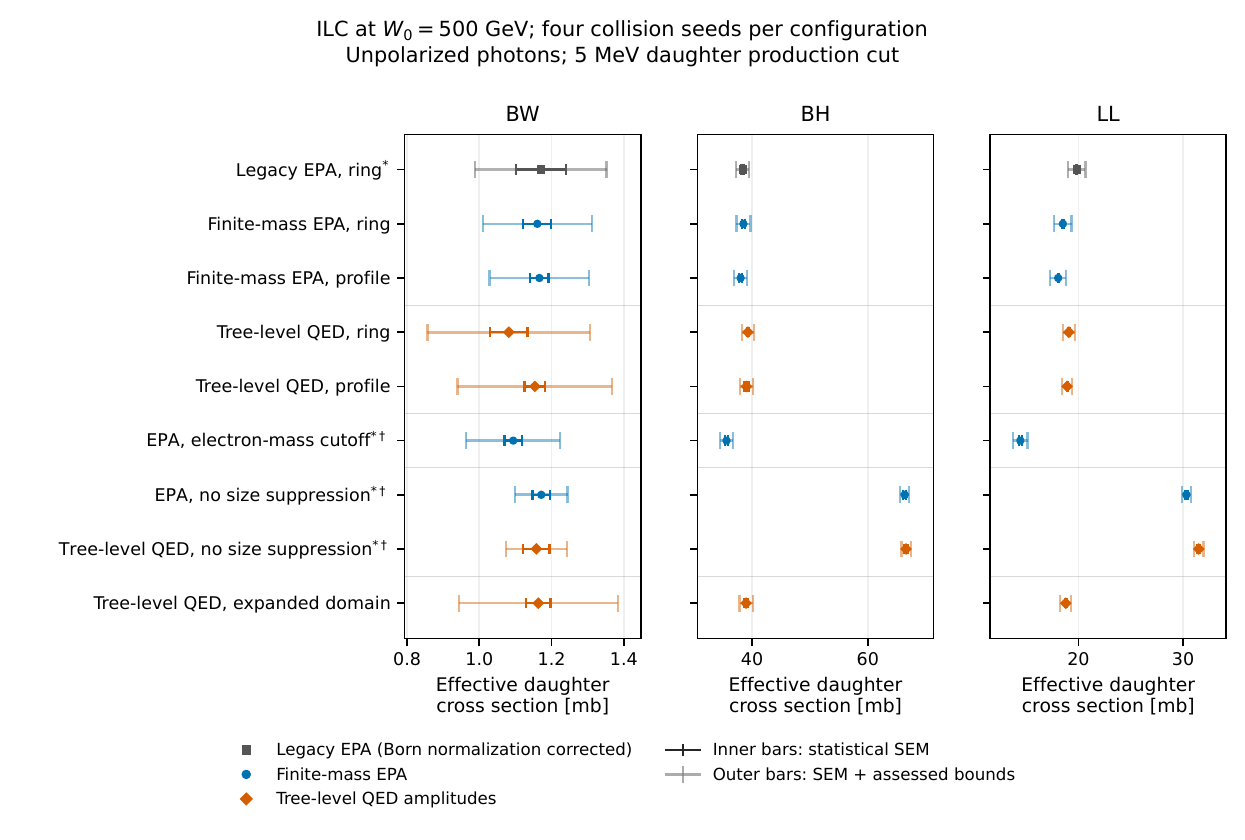}
\caption{Central effective daughter production cross sections for nine
ILC 500 GeV configurations, each using four matched collision runs. Each
point provides the luminosity weighted daughter yield. Uncertainty is evaluated from statistics and from assessed bounds across multiple seeds and configurations.
Superscripts $^{*}$ and $^{\dagger}$ identify a corresponding prescription
or the marked option documented for \GP\ and \CAIN,
respectively~\cite{pairs,cain244}.}
\label{fig:pair-controls}
\end{figure}

\paragraph{Controlled beam pair process comparison}
We separate the implementation of beam pair processes into nine configurations,
each using four matched collision seeds. All retain the
5 MeV daughter production threshold. Figure~\ref{fig:pair-controls}
shows all three luminosity weighted channels.

For LL, the refined EPA treatment changes the result by $-1.321\pm0.089$ mb, or $-6.66\%$. 
This combines the finite-mass flux, angular sampling and coupling constant changes. Replacing ring placement with the impact parameter profile changes EPA by $-0.450\pm0.081$ mb.
The full hard-process treatment then changes the profile result by
$+0.852\pm0.046$ mb. These shifts depend on the starting model: the
difference between the EPA and full-amplitude profile changes is
$0.283\pm0.081$ mb.

The full amplitude result is $19.087\pm0.103$ mb, 0.59\% below
the published \GP\ value. With beam-size suppression disabled, it becomes
$31.502\pm0.108$ mb, 0.94\% below the independent BDK calculation's
31.8 mb in Ref.~\cite{pairs}. BDK includes electroweak diagrams beyond
the present QED scope. The exact calculation also uses a minimum virtuality
spectator assignment to define the produced daughters in identical lepton
final states, which is distinct from the EPA construction.

Introducing the electron mass EPA cutoff lowers LL by $4.063\pm0.093$ mb. Doubling the transverse domain and cell counts at fixed spacing changes
the full-profile LL result by $-0.123\pm0.081$ mb, or $-0.65\%$.
That control includes both field and source coverage and virtual photon
acceptance. No additional LL kernel defect is identified by these checks.
The remaining percent level difference is compatible with the measured
model and domain sensitivities.

\FloatBarrier

At 1 TeV, the reconstruction of the ILC Technical Design Report's
B1b upgrade parameter set gives a ratio of full and peak spectrum distances relative to the Forward Calorimetry (FCAL) collaboration's distribution of
$0.0259$ and $0.0182$ respectively~\cite{fcal}. The
corresponding conditional first Wasserstein distances are
$2.51/0.160~\GeV$ for the mean of 12 independent collisions. Broad tail and
peak shapes are similar, but the reference input and its correlations are not reconstructed. We suspect that the reference simulations have accelerator and chirp effects, which are not present in the \FeeBL\ comparison, which could explain the slightly sharper peak and smaller tail.
\begin{figure}[htbp]
\centering\includegraphics[width=\textwidth]{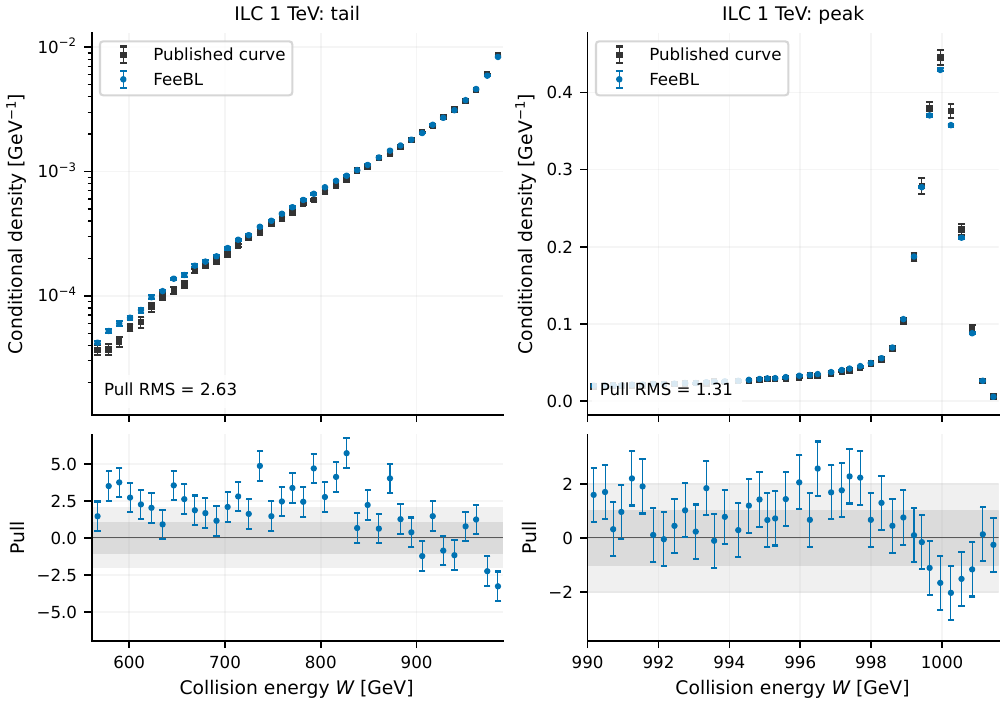}
\caption{Tail and peak ILC 1 TeV luminosity spectra compared with
Ref.~\cite{fcal}. The left panel shows 38 tail bins from 560.8 to 990 GeV and excludes the
peak region shown in the right panel, which uses 40 bins from 990 to
1001.6 GeV.}
\label{fig:ilc-spectrum}
\end{figure}

\subsection{HALHF luminosity spectra and beam pairs at 250 GeV}
The September 2026 HALHF parameters~\cite{halhf} are in a slightly different regime as they have asymmetric beams and energies and a stronger-field. High-energy fractions
in Table~\ref{tab:halhf} use incoming energies.

\begin{table}[htbp]
\centering\small
\caption{Historical HALHF luminosity in $10^{34}\,\mathrm{cm}^{-2}\mathrm{s}^{-1}$,
in fraction $f_{0.99}$. Each result has one seed.}
\label{tab:halhf}
\begin{tabular}{@{}rrrrrrrr@{}}
\toprule
$W_0$ [GeV]&$\mathcal L_{\rm ref}$&$\mathcal L$&Difference&$f_{0.99,\rm ref}$&$f_{0.99}$&$D_{\rm full}$&$D_{\rm peak}$\\
\midrule
250&1.14&1.11208&$-2.45\%$&0.66&0.65212&0.00940&0.01442\\
380&1.75&1.71061&$-2.25\%$&0.57&0.55784&0.01151&0.01599\\
550&2.55&2.46365&$-3.39\%$&0.50&0.48812&0.01281&0.01223\\
\bottomrule
\end{tabular}
\end{table}
The agreement is plausible at the few-percent luminosity level, and given that the original run settings were not found, this is within the level that can be explained by methodological differences in the simulations.

\begin{figure}[htbp]
\centering\includegraphics[width=\textwidth]{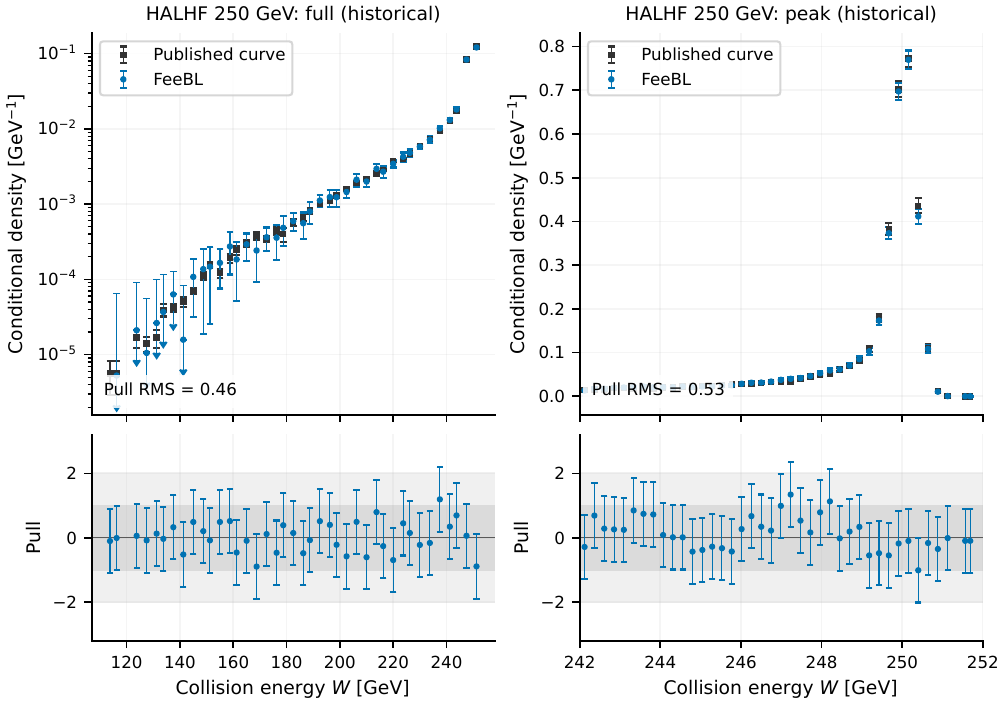}
\caption{HALHF 250 GeV full and peak spectra relative to
Ref.~\cite{halhf}. Both panels use 40 display bins, with a 242-252 GeV peak axis.
The source and simulated curves agree well.}
\label{fig:halhf-spectrum}
\end{figure}

Figure~\ref{fig:halhf-spectrum} displays the broad shape agreement at
250 GeV. As such, this gives further evidence that differences seen in Table~\ref{tab:halhf} are methodological and not physical. 

\begin{figure}[p]
\centering\includegraphics[width=\textwidth]{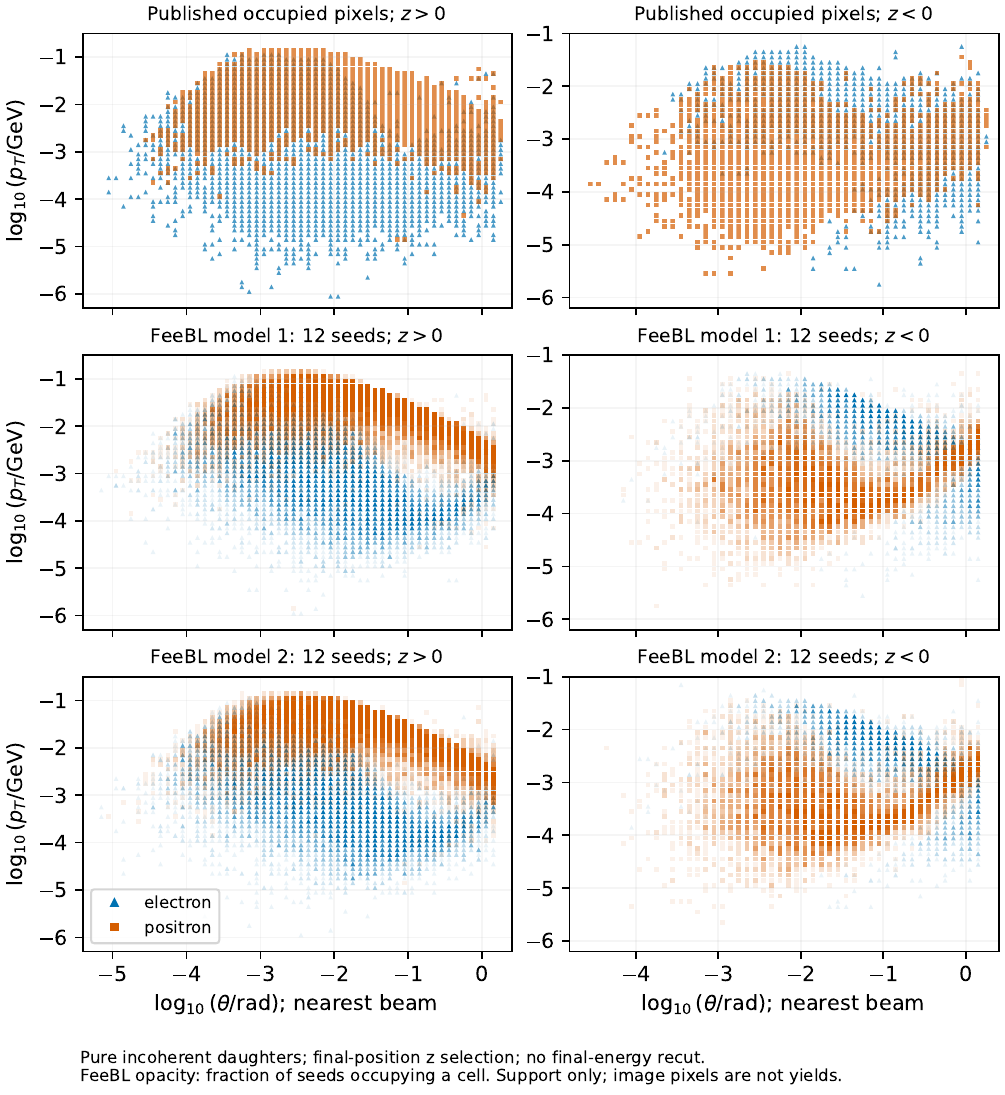}
\caption{Polar angle and transverse momentum of incoherent pairs for HALHF at 250 GeV. The top row shows the published occupied
cells~\cite{halhf}; the middle and bottom rows show \FeeBL\ pair modes
1 and 2. Columns select final position $z>0$ and $z<0$; the angle is
folded toward the nearer beam axis. Color and marker shape distinguish
electrons and positrons. In simulated panels, opacity is the fraction
of twelve independent stored samples occupying a cell while, in the reference panels, no such fraction could be computed so it merely reflects presence of a particle.}
\label{fig:halhf-pair-phase-space}
\end{figure}

Figure~\ref{fig:halhf-pair-phase-space} isolates incoherent daughters
in both pair models, using twelve independent
collisions per model. Final position selections $z>0$ and $z<0$ follow the
reference panel labels.
The reference does not fully specify its algorithm, so this comparison remains qualitative and not quantitative. Reference image cells indicate occupied cells rather than event counts or a normalized density. The simulated plots use a density computed from the fraction of independent stored samples occupying each cell.

\FloatBarrier
\subsection{CPU and GPU comparison}
\label{sec:backend-comparison}
We compare eight CPU and GPU runs with different seeds using FCC-ee inputs from
Ref.~\cite{fcccollective2025}: $120\GeV$ per beam, 30 mrad full crossing
angle and 50\% crab-waist strength. The FCC-ee inputs are particularly useful for a stress and consistency test as the long bunch length and narrow transverse RMS spread, coupled with the crab-waist scheme, resulted in FCC-ee simulations being slowest and greediest in terms of resources needed to achieve numerical convergence. 
We use the published equilibrium RMS moments for the beams; ring equilibrium is not calculated nor currently supported. 
Each beam starts with $10^6$ macroparticles and cut at three RMS bunch lengths. 
The mesh has $512\times1024$ transverse cells, 64 slices and two substeps.
We disable spin and beam pairs for these runs and examine only the luminosity spectra.

For smaller scale local tests, such as most of the tests in this paper, run time was observed as 2.5 $\pm$ 1.1 minutes, while being run on an average home computer. The stress testing done here was done on DESY Maxwell using two Intel Xeon Silver 4216 processors and two 32 GiB NVIDIA V100 SXM2 GPUs. 
Two independent collisions run concurrently, each on 16 physical cores and, for GPU execution, one GPU. Both tests use the same executable, built with GCC 12.2 and CUDA 12.9. The full allocation, including build and trials, was 32 minutes and 28 seconds.

Table~\ref{tab:cpu-gpu-time} presents the timing results. The median time ratio is 1.82 with a range of 1.79 to 1.86. 
Host physics still occupies 44.45 seconds per GPU assisted collision and thus approximately 68\% of elapsed time. This result means that further optimizations are needed before \FeeBL~is limited by GPU saturation and, therefore, we expect future development will improve this performance further.

\begin{table}[htbp]
\centering\small
\caption{FCC-ee stress test timing results. Collision time is the mean from eight seeds; other rows sum eight collisions or four concurrent two collision batches.
CPU consumption is user plus system time. Assigned GPU time includes host work and idle periods. 
CUDA intervals include cuFFT but are not total collision times.}
\label{tab:cpu-gpu-time}
\begin{tabularx}{\textwidth}{Yrr}
\toprule
Time on stage and on device [s]&CPU&GPU with host threads\\
\midrule
Mean elapsed collision time&$118.46 \pm 0.67$&$65.106 \pm 0.069$\\
Elapsed time per two collision batch&$119.39 \pm 0.66$ & $65.65 \pm 0.29$\\
CPU core time per collision&$1045.1 \pm 8.5$ & $237.46 \pm 0.27$\\
CUDA kernel and cuFFT event intervals per collision&---&$6.42634 \pm 0.00032$\\
Host and device transfer event intervals per collision&---&$0.79595 \pm 0.00086$\\
GPU device time per collision&---&$65.106 \pm 0.069$\\
\bottomrule
\end{tabularx}
\end{table}

\begin{figure}[p]
\centering\includegraphics[width=\textwidth]{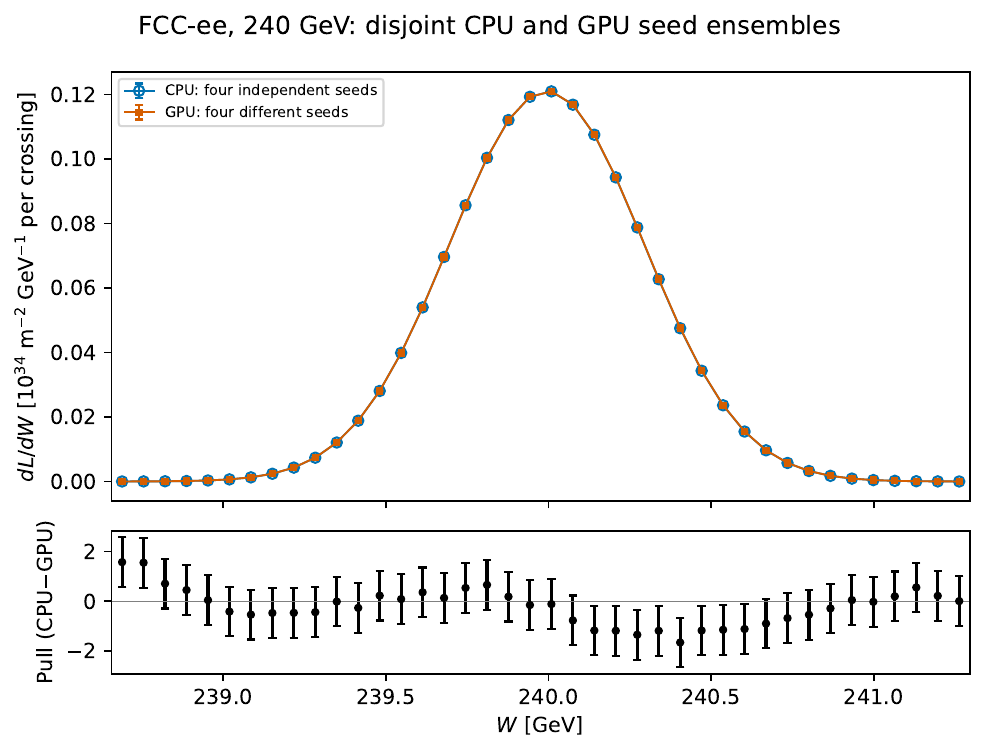}
\caption{FCC-ee 240~GeV comparison of independent ensembles of four runs, using one
million macroparticles per beam and a $512\times1024\times64$
cell mesh. 
The displayed energy interval has forty 66 MeV bins. Blue open
circles show the 16-thread CPU result and orange squares show the result from one V100 with 16 host threads. The CPU seeds are 72942001, 003, 005 and 007 while the GPU seeds are 72942002, 004, 006 and 008. 
Errors are standard errors across complete collisions. Pulls are CPU minus GPU divided by the combined standard error.}
\label{fig:cpu-gpu-spectrum}
\end{figure}

Across all eight seeds the mean luminosity per crossing is $(9.1143\pm0.0012)\times10^{32}\,\mathrm{m}^{-2}$. Each matched histogram is identical at six-significant-digit
output precision; this paired closure is retained separately from
Fig.~\ref{fig:cpu-gpu-spectrum}. 
That figure uses four odd-seed CPU runs and four even-seed GPU runs, with no shared seeds.
Its difference covariance is the sum of the two covariance matrices of the mean, without a paired cross term. The displayed subtraction is CPU minus GPU, with largest absolute pull 1.66. The timing table retains all eight runs per backend.

\FloatBarrier

\section{Depolarization and spin tests}
\label{sec:spin}
For investigating spin physics we use the ILC 2006 ``nominal'' settings and results~\cite{spin2006}. These results report a luminosity weighted loss of
$0.10\%$, comprising $0.08\%$ from precession and $0.02\%$ from spin
flips. It does not specify an energy cut or an explicit
per-beam versus combined formula. The later implementation paper gives approximately
$0.23\%$ for \GP++ and $0.24\%$ for \CAIN~\cite{spin2008}.
These are different historical comparisons, not interchangeable estimates
to be pooled without their conditions. The earlier table also gives
$0.19\%$ for Large Y and $0.24\%$ for Low P; the later code values
are $0.46$--$0.57\%$ and $0.41$--$0.50\%$, respectively. The repeated
change of scale motivates examining historical definitions and model
settings rather than assigning the nominal difference to one implementation
defect.

\begin{table}[htbp]
\centering\small
\caption{Published nominal ILC predictions for initially fully polarized
beams. Values are absolute losses expressed in \%.}
\begin{tabular}{@{}lrrr@{}}
\toprule
Publication/code&$\beta_x$ [mm]&Precession [\%]&Total [\%]\\
\midrule
EPAC 2006, \CAIN~\cite{spin2006}&21&0.08&0.10\\
Rimbault et al. 2009, \GP++~\cite{spin2008}&20&0.17&0.23\\
Rimbault et al. 2009, \CAIN~\cite{spin2008}&20&0.19&0.24\\
PAC 2011, \CAIN~\cite{spin2011}&20&0.17&0.22\\
\bottomrule
\end{tabular}
\end{table}

The 2009 conclusion quotes $0.23\pm0.01\%$ without defining a confidence
level. The 2011 study estimates an absolute statistical uncertainty near
$0.10\%$. The EPAC 2008 result $0.21\pm0.01\pm0.03\%$ instead expresses
model sensitivities to field and spin-flip energy-ratio variations
\cite{spin2008epac}. Verifying the inconsistency in these results is beyond the scope of a demonstration so we note that the reference material itself is uncertain within $0.1\%$ or so, and thus an agreement within this interval is treated as plausible. 

The 125 GeV-per-beam table in Yokoya's presentation~\cite{yokoya2022}
separately quotes electron and positron depolarization, but its weighting,
spin projection, and normalization must be resolved before a numerical
comparison with the collision-weighted loss here.

\subsection{Spin equation and corrections tested independently}
Classical precession follows the Thomas-Bargmann-Michel-Telegdi equation~\cite{bmt}. Beamstrahlung spin transitions and conditional no-emission evolution follow the stated \CAIN\ radiation model~\cite{cain21,cain244}.
Independent adaptive Bessel-function quadrature tests the transition law.
The quantum parameters are $\chi=0.01,0.05,0.2,1,3$, with two proposal
exposures and four initial spin
ensembles. Forty settings each contain two million trials, giving 200
probability, radiated-energy and spin means plus 40 norm/energy checks.
A retrospective within-family assessment uses a fixed 5\% simultaneous
empirical Bernstein criterion~\cite{maurer2009}. All 200 means satisfy
the criterion; the largest absolute discrepancy is 0.706 times its
simultaneous sampling radius. Finite-step and numerical allowances remain
separate, and all 40 bounds checks report zero invalid trials. This
result assumes independent bounded trials within each setting and does not require independence between settings or establish the absence of smaller biases. 

An independent
oblique-field BMT calculation agrees to $1.1\times10^{-16}$ in spin
components. Four weak-field checks recover longitudinal/transverse damping
$0.777716$ versus $7/9$, and equilibrium polarization $0.923734$ versus
$8/(5\sqrt3)$. Five anomalous-moment checks test the implemented
approximation and asymptotes~\cite{esberg}.
\begin{figure}[htbp]
\centering\includegraphics[width=\textwidth]{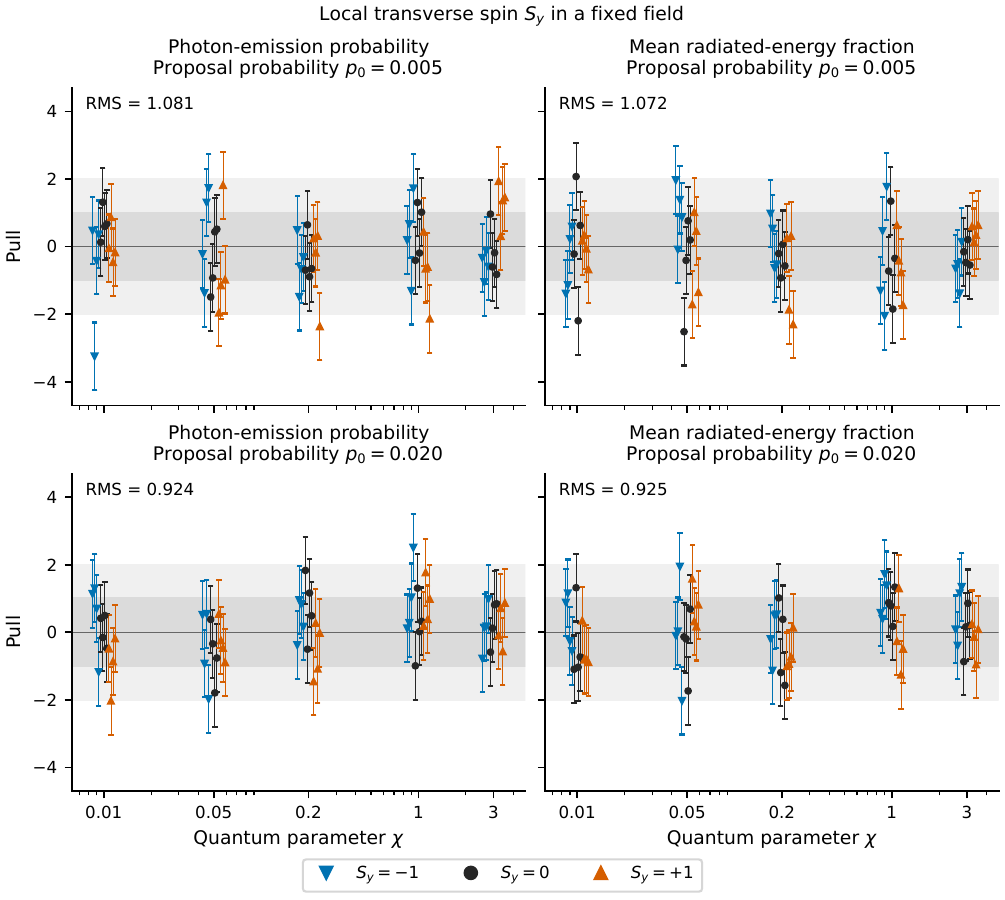}
\caption{Sampling pulls for photon emission probability and mean radiated
energy fraction relative to independent deterministic quadrature.
Here $p_0$ is the dimensionless probability of proposing a photon in each trial, before quantum and spin-dependent acceptance; it is not the final photon-emission probability.
There are 120 independent runs: three initial transverse spin projections,
five quantum parameters, two proposal probabilities and four fresh seeds
per setting, each with two million trials.}
\label{fig:sampler}
\end{figure}
Figure~\ref{fig:sampler} uses an independent-seed replication rather than
reusing one seed across settings. Its photon probability and radiated energy
pull RMS values are 1.0053 and 1.0012. For initial transverse spin component, $S_y=+1$, the mean
pulls across 40 independent runs are $-0.116\pm0.169$ for photon
probability and $-0.232\pm0.141$ for radiated energy. Here the errors are
empirical SEMs across the fixed quantum parameter and exposure settings.
The latter is the largest subset offset, at 1.65 SEMs.

\subsection{Inclusive and peak depolarization samples}
At the original 10k particle settings, corrected per-beam helicity losses
are $0.21350$ and $0.22365\%$. The fixed-axis spin-product loss is
$0.21907\%$, versus $0.21846\%$ for local helicity. Replacing the
correlated fixed-axis product with separate mean projections changes the
loss by only $0.000113\%$ in this case.

The energy selections in Table~\ref{tab:cuts} are at $W_0=500\GeV$. In the peak region, above 98\% of
nominal energy, the mean local spin-product loss is 31\% smaller than its
inclusive value; above 99\%, it is 35\% smaller.

\begin{table}[htbp]
\centering\small
\caption{Three-seed energy-selected results with 40k macroparticles
per beam at fixed coarse mesh resolution. Losses are in \% relative to
unit helicity. Errors are SEMs of the seed mean only.}
\label{tab:cuts}
\begin{tabular}{@{}lrrrr@{}}
\toprule
Selection&Retained $\mathcal L$ [\%]&$\Delta P_-$&$\Delta P_+$&$\Delta P_{\rm pair}$\\
\midrule
Inclusive&100&$0.2228\pm0.0018$&$0.2218\pm0.0037$&$0.2222\pm0.0018$\\
$W\geq0.98W_0$&$83.40\pm0.11$&$0.1534\pm0.0015$&$0.1536\pm0.0022$&$0.1534\pm0.0009$\\
$W\geq0.99W_0$&$74.19\pm0.10$&$0.1442\pm0.0016$&$0.1443\pm0.0021$&$0.1442\pm0.0009$\\
\bottomrule
\end{tabular}
\end{table}

\subsection{Convergence test of spin tracking}
Three independent 40k particle seeds give the inclusive fixed-axis loss
$0.22282\pm0.00183\%$, with seed standard deviation $0.00318\%$.  For three independent normal seed estimates, a 95\% Student-$t$ interval gives approximately $[0.21495,0.23069]\%$. Disabling spin flips gives $0.16752\pm0.00105\%$ and the paired difference
$0.05530\pm0.00079\%$.

\begin{table}[htbp]
\centering\small
\caption{Matched-seed numerical variations at $\beta_x=21$ mm. $N_m$ is
macroparticles per beam, $N_z$ slices and $N_t$ encounter subdivisions.
Shifts between rows are numerical sensitivities, not independent errors.}
\begin{tabular}{@{}lrrrrr@{}}
\toprule
Variation&$N_m$&$N_x=N_y$&$N_z$&$N_t$&$\Delta P_{z,L}$ [\%]\\
\midrule
Baseline&10\,000&128&32&2&0.21907\\
Particles&20\,000&128&32&2&0.22245\\
Particles&40\,000&128&32&2&0.21948\\
Transverse cells&40\,000&256&32&2&0.22571\\
Slices&40\,000&128&64&2&0.22457\\
Time subdivisions&40\,000&128&32&4&0.22476\\
Combined&40\,000&256&64&4&0.23518\\
\bottomrule
\end{tabular}
\end{table}
The combined refinement increases the matched loss by 7.15\%, exceeding
the seed SEM. Numerical convergence is therefore unresolved even though the
transport controls pass. Changing only $\beta_x$ from 21 to 20 mm gives
$0.22985\%$, a 4.72\% increase, consistent with the 4.85\% sensitivity
of the approximate Yokoya and Chen precession estimate~\cite{yokoya}.
This small optical change is still not sufficient to explain the differences between reference simulations and \FeeBL\ simulations.

Table~\ref{tab:polarization-inputs} specifies the additional 2009 optical
configurations. Their evaluated transverse sizes use the published
emittances and beta functions.
\begin{table}[htbp]
\centering\small
\caption{Inputs for the 2009 spin comparisons~\cite[Table~1]{spin2008}.
Each beam has energy $250\GeV$, zero intrinsic energy spread and unit initial
longitudinal polarization. The crossing is head-on. Emittances are in
mm\,mrad, beta functions in mm, transverse sizes in nm and bunch length
in $\mu$m. All use $128^2$ transverse cells, $N_z=32$ and $N_t=2$.
The Low N sample has fewer macroparticles and requires separate refinement.}
\label{tab:polarization-inputs}
\begin{tabular}{@{}lrrrrrr@{}}
\toprule
Case & $N/10^{10}$ & $\epsilon_{nx}/\epsilon_{ny}$ & $\beta_x/\beta_y$ & $\sigma_x/\sigma_y$ & $\sigma_z$ & $N_m$\\
\midrule
Nominal & 2 & 10/0.04 & 20/0.4 & 639.4/5.72 & 300 & 40,000\\
Low N & 1 & 10/0.03 & 11/0.2 & 474.2/3.50 & 200 & 10,000\\
Large Y & 2 & 10/0.08 & 11/0.6 & 474.2/9.91 & 500 & 40,000\\
Low P & 2 & 10/0.036 & 11/0.2 & 474.2/3.84 & 200 & 40,000\\
\bottomrule
\end{tabular}

\end{table}

These additional configurations provide a more informative
comparison. Their reported beam emittances and beta functions set
the sizes; reconstructed Gaussian beams, zero energy spread and IP waists
are used where the original particle records are absent.
\begin{table}[htbp]
\centering\small
\caption{Inclusive fixed-axis spin-product losses in \% for the 2009
configurations~\cite{spin2008}. Parentheses give the primary macroparticle
count per beam.}
\begin{tabular}{@{}lrrrr@{}}
\toprule
Optics & GUINEA-PIG++ [\%] & CAIN [\%] & FeeBL [\%] & $N_m$\\
\midrule
nominal & 0.23 & 0.24 & 0.2298 & 40000\\
low n & 0.13 & 0.13 & 0.1220 & 10000\\
large y & 0.46 & 0.57 & 0.5073 & 40000\\
low p & 0.41 & 0.50 & 0.4897 & 40000\\
\bottomrule
\end{tabular}

\end{table}
Large Y and Low P lie between the published code values and are assessed
as plausible. Their 10k particle losses, 0.49964 and 0.51113\%, change
by $+1.54\%$ and $-4.19\%$ when the particle count is increased, restoring
the published ordering. Nominal rounds to 0.23\% at the reference precision.
Low N is 6.18\% below both references and is thus plausible in scale,
but is not within their spread.

\begin{figure}[htbp]
\centering\includegraphics[width=\textwidth]{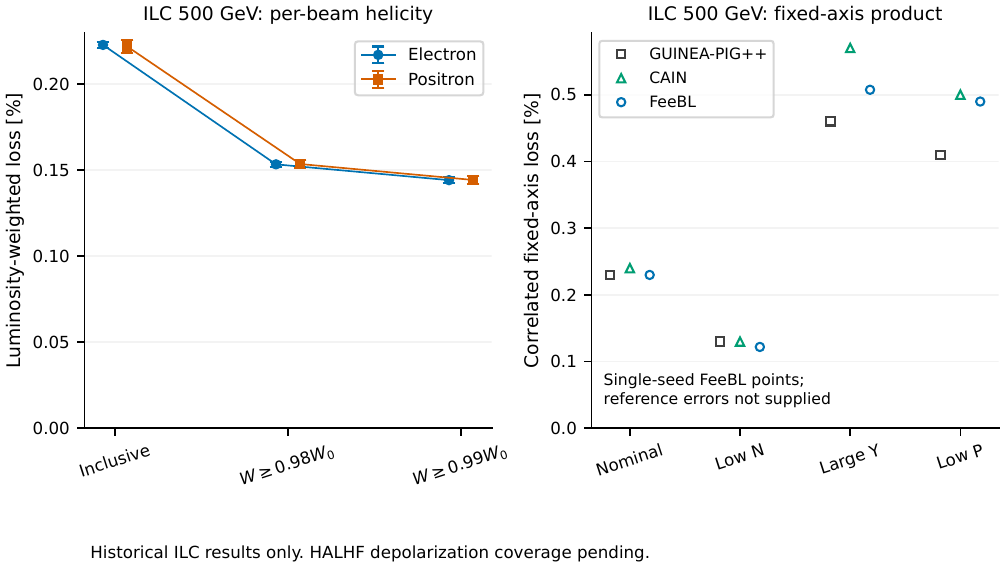}
\caption{Depolarization at ILC with collision energy of 500 GeV. Left: luminosity-weighted helicity
loss for each initially fully polarized beam, inclusively and with the
98\% and 99\% collision-energy selections. Capped errors are three seed
SEMs at fixed numerical settings. Right:
inclusive correlated fixed axis spin-product losses for the four 2009
configurations~\cite{spin2008}. These retained \FeeBL\ points
have one seed each and, as such, errors have not been measured.}
\label{fig:depolarization-scenarios}
\end{figure}

Figure~\ref{fig:depolarization-scenarios} displays the per-beam selection
dependence and the configuration comparison separately. The available
independent seed per-beam results cover ILC. Extending this comparison to
HALHF requires dedicated spin-enabled ensembles and results for HALHF, which needs to be addressed in future studies of HALHF. 
The current code rejects spin tracking with the canonical crab-waist configuration, so FCC-ee depolarization is not included. Support for polarized beams with crab-waist optics is planned for future development.

\FloatBarrier

\section{Monochromatization with horizontal dispersion}
\label{sec:mono}
Direct resonant Higgs production via $e^+e^-\to H$ requires collision energies near
125 GeV, rather than the typical 240-250 GeV for Higgsstrahlung. The Higgs inclusive width of $\approx$4.1 MeV thus requires a machine with a peak width of comparable or smaller width, a feat that is well beyond what was demonstrated at the LEP Z pole line scan~\cite{higgsresonance2022,higgslineshape2016}. 
Narrowing alone does not establish a useful measurement, as luminosity, ISR,
calibration, acceptance, and backgrounds determine the available information. A direct Yukawa coupling measurement for the electron is a valuable test of Standard Model physics. New beam delivery techniques and new ways to take advantage of beam-beam physics to narrow the width of the collision energy distribution are therefore a worthwhile endeavor. Such studies are referred to as monochromatization due to the emphasis on reducing the collision energy distribution towards one energy.

Currently \FeeBL\ only supports horizontal dispersion for monochromatization so we restrict this study to that case. Horizontal dispersion correlates
particle energy with transverse position so that spatial overlap preferentially
selects compensating energy deviations~\cite{monoprinciple}. It need not
reduce either incoming beam's energy spread.

\subsection{Derivations and definitions for monochromatization}
Horizontal dispersion correlates a particle's energy with its position at
the interaction point. Opposite correlations in the two beams favor
collisions between particles with compensating energy deviations. The
Gaussian overlap calculation is established accelerator physics
\cite{bogomyagkov2017,monoprinciple}; here we retain unequal beam sizes and
energy spreads to cover asymmetric collisions as well.

For beam $i$, define the fractional energy deviation as done in Sec.~\ref{ssec:mono} and with a standard deviation of $s_i$.
Write $x_i=x_{\beta i}+D_i\delta_i$, where $D_i$
is the horizontal dispersion and $b_i$ is the RMS betatron size, excluding
the dispersive contribution. Thus, $D_i$ and $b_i$ are lengths, and
$B_x=b_1^2+b_2^2$ has units of length squared. Both dispersions refer to
the same horizontal coordinate at the collision point. The incoming
$\delta_i$ and $x_{\beta i}$ are taken to be independent Gaussian
variables. The collisions are treated as head-on and ultrarelativistic.

Integrating the horizontal profiles gives the overlap factor
$\exp[-(D_1\delta_1-D_2\delta_2)^2/(2B_x)]$.
The resulting weighted variance of $\delta_1+\delta_2$ is
\begin{equation}
 V_{\rm coll}=
 \frac{(s_1^2+s_2^2)B_x+s_1^2s_2^2(D_1+D_2)^2}
 {B_x+D_1^2s_1^2+D_2^2s_2^2}.
 \label{eq:lambda}
\end{equation}
The numerator and denominator both have units of length squared, so
$V_{\rm coll}$ is dimensionless. To first order in the energy deviations,
\begin{equation}
 \sigma_{W,\rm geom}=\frac{W_0}{2}\sqrt{V_{\rm coll}},
 \qquad
 \Lambda_{\rm ideal}=\sqrt{\frac{s_1^2+s_2^2}{V_{\rm coll}}},
 \qquad
 \frac{\mathcal L(D_1,D_2)}{\mathcal L(0,0)}
 =\sqrt{\frac{B_x}{B_x+D_1^2s_1^2+D_2^2s_2^2}}.
 \label{eq:mono-width-luminosity}
\end{equation}
Here the luminosity ratio holds at fixed bunch populations, collision
rate and betatron optics. For $D_1=-D_2=D$, it equals
$1/\Lambda_{\rm ideal}$, including unequal beams. For equal beams,
$s_1=s_2=s$ and $b_1=b_2=b$, the familiar result is
$\Lambda_{\rm ideal}=\sqrt{1+D^2s^2/b^2}$
\cite{bogomyagkov2017,monoprinciple}. Equal same-sign dispersions leave
$V_{\rm coll}$ unchanged when $s_1=s_2$.

We let $f(W)$ denote a normalized luminosity spectrum. Its fitted model $p_W(W)$ provides
a separate definition of the peak width for radiative spectra. We use
Wilson's convolution of a beta distribution with an endpoint Gaussian~\cite{wilson2023}, following the beta-plus-endpoint description
used in CIRCE~\cite{ohl1997circe}. To make the units explicit, write the
model directly as a density in collision energy:
\begin{align}
 p_W(W)&=f_{\rm peak}\,\phi(W-W_\star;\sigma_G)
 +\frac{1-f_{\rm peak}}{\mathrm B(\alpha,\beta)}
 \int_0^1 x^{\alpha-1}(1-x)^{\beta-1}
       \phi(W-xW_\star;\sigma_G)\,dx,
 \label{eq:betagaussian}\\
 \phi(y;\sigma_G)&=\frac{1}{\sqrt{2\pi}\sigma_G}
                  \exp\!\left[-\frac{y^2}{2\sigma_G^2}\right].
 \nonumber
\end{align}
Here $W_\star$ is the fitted endpoint energy before convolution,
$\alpha,\beta>0$ are dimensionless tail parameters,
$\mathrm B$ is the Euler beta function, and $0\leq f_{\rm peak}\leq1$.
Both $W_\star$ and $\sigma_G$ have energy units; $p_W$ and $\phi$ have
inverse-energy units. This is Wilson's model with
$\sigma_G=W_\star\sigma$, where his $\sigma$ is a fractional Gaussian
width; no additional Jacobian is required in this density form.

We report $\sigma_G$ as the fitted Gaussian response width.
The corresponding Gaussian FWHM is
$2\sqrt{2\ln2}\,\sigma_G$; it need not equal the FWHM of the complete
tail-plus-peak distribution. The model is empirical, and thus identifying its Gaussian width does not prove that every contribution to the simulated peak is Gaussian.

The model is integrated over the histogram bins and fitted by density
least squares with nonnegative component amplitudes. Statistical
uncertainties are evaluated across independent simulation seeds. The
measured narrowing factor compares fitted widths with and without
dispersion at otherwise matched settings.

\subsection{Linear collider monochromatization with conserved luminosity}
The ILC/LCF and HALHF families retain published 250 GeV normalized
emittances, populations, relative energy spreads, lengths, and rates while
changing energies~\cite{ilcstaging2017,halhf}. Table~\ref{tab:inputs125}
defines these exploratory assumptions. The LCF interpretation doubles only the crossing rate
\cite{lcf2025}, so we treat it as having the same per-crossing spectrum as ILC.

\begin{table}[htbp]
\centering\small
\caption{125 GeV linear-collider input families. Ordered entries identify
electron and positron; a single value applies to both. Luminosities are per IP.}
\label{tab:inputs125}
\begin{tabularx}{\textwidth}{@{}Ycc@{}}
\toprule
Quantity&ILC / LCF&HALHF\\
\midrule
Beam energy [GeV]&62.501531&187.5, 20.833333\\
Population [$10^{10}$]&2&1, 3\\
Normalized $\epsilon_{n,x}$ [$\mu$m]&5&90, 10\\
Normalized $\epsilon_{n,y}$ [nm]&35&320, 35\\
Reference $\beta_x^*,\beta_y^*$ [mm]&13, 0.41&3.3, 0.10\\
Bunch length [$\mu$m]&300&150\\
Relative energy spread [\%]&0.188, 0.150&0.150\\
Full crossing angle [mrad]&14&0 (assumed)\\
Ideal RF-like tilt&on&off\\
Vertical waist displacement [$\mu$m]&0&75\\
Crossings/s&6560 / 13120&16000\\
\bottomrule
\end{tabularx}
\end{table}

The ideal ILC crab map supplies perfectly prepared spatial tilt with its
canonical energy companion. It does not model RF curvature, phase noise or
cavity transport. Physical crossing still enters the invariant mass. The
HALHF head-on geometry and 9:1 energy asymmetry are explicit assumptions.

The initial scan contains sixteen linear-collider discovery points, varying
horizontal focusing and dispersion with matched zero and same-sign controls.
The selected points reduce $\beta_x$ by four and apply
$(D_1,D_2)=(+0.3712,-0.3712)$ mm for ILC/LCF or
$(+0.5194,-0.5194)$ mm for HALHF. Vertical beta remains unchanged.
The geometric expectations are a factor of two narrowing and unchanged
luminosity.

\begin{figure}[htbp]
\centering\includegraphics[width=\textwidth]{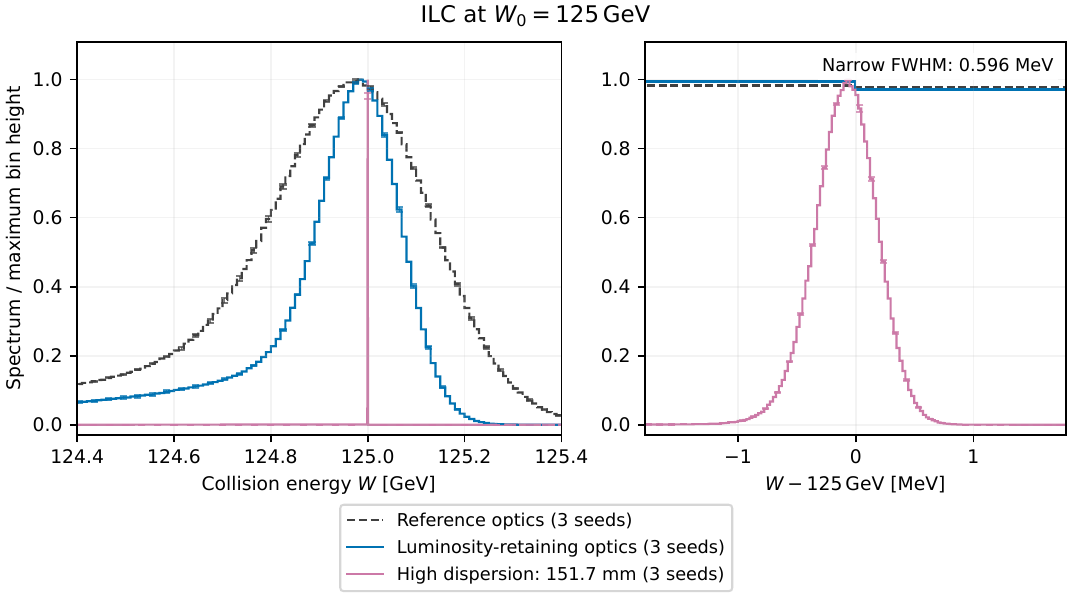}
\caption{ILC spectra at 125 GeV for the reference, luminosity conserving quarter-beta opposite-dispersion, and high dispersion settings. Each curve is normalized to the maximum bin height so that shapes can be easily compared.}
\label{fig:ilcmono}
\end{figure}

\begin{figure}[htbp]
\centering\includegraphics[width=\textwidth]{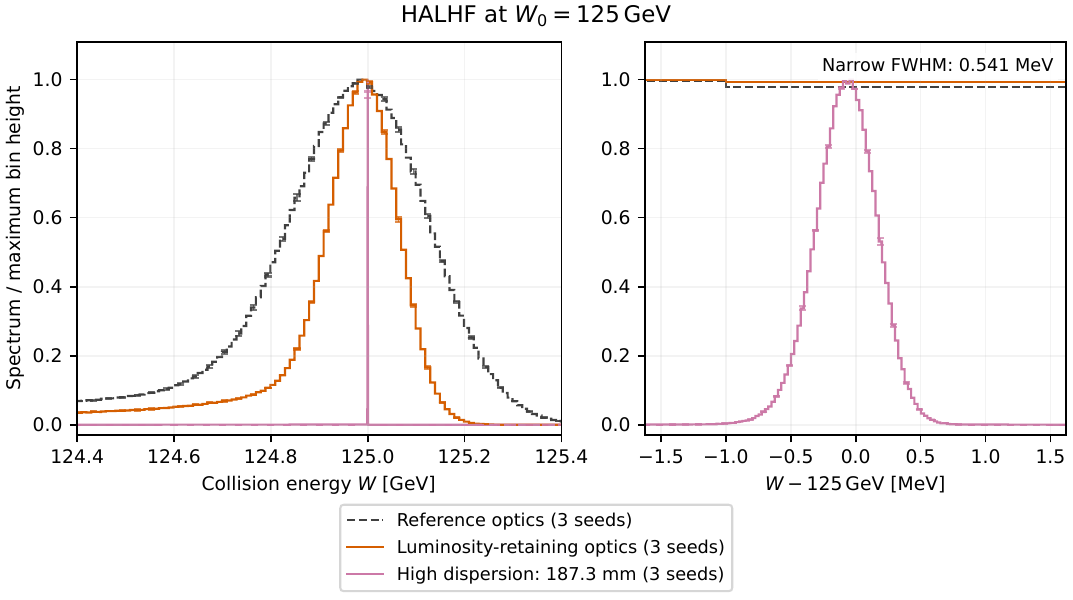}
\caption{HALHF confirmation spectra with fixed 9:1 energy ratio and
assumed head-on geometry. Settings and normalization conventions follow
Fig.~\ref{fig:ilcmono}.}
\label{fig:halhfmono}
\end{figure}

\begin{table}[htbp]
\centering\small
\caption{Independent confirmation of the selected 125 GeV points. Errors
are SEMs across three seed pairs.}
\begin{tabular}{@{}lrr@{}}
\toprule
Quantity & ILC/LCF & HALHF\\
\midrule
Baseline $\sigma_G$ [MeV] & $150.1\pm1.1$ & $133.8\pm1.1$\\
Selected $\sigma_G$ [MeV] & $76.50\pm0.35$ & $67.21\pm0.31$\\
Paired width ratio & $0.5096\pm0.0015$ & $0.5023\pm0.0019$\\
Paired luminosity ratio & $1.023\pm0.023$ & $1.0010\pm0.0031$\\
\bottomrule
\end{tabular}

\end{table}
All twelve confirmation fits pass their stated response checks. Approximate
95\% Student-$t$ intervals for luminosity retention are [0.9241,1.122] and
[0.9875,1.015]. This result indicates that monochromatization, even at moderate values of horizontal dispersion, is a promising direction for improved collision energy spread at linear colliders. With a future feasibility study and technology demonstration this could be envisioned as an upgrade path for a linear collider. This is not a claim that such a dispersion could be introduced without loss of performance in other things, it is merely a demonstration that the beam-beam physics support the potential of this kind of upgrade. Figures~\ref{fig:ilcmono} and \ref{fig:halhfmono} show the redistribution
into the peak under the selected settings and their matched baselines.


\subsection{High dispersion without a luminosity floor}
To investigate resonances as narrow or narrower than the Higgs width, we extend
the opposite-sign horizontal dispersion scan at $W_0=125\GeV$ without
requiring luminosity conservation. The horizontal beta functions remain
$\beta_x^*=3.25$ mm for ILC/LCF and $0.825$ mm for HALHF. The
zero dispersion controls use these same beta functions. ILC retains its
14 mrad crossing angle with ideal spatial crab compensation; HALHF uses
the head-on configuration with a $9{:}1$ beam-energy ratio. Table~\ref{tab:highd} reports the observed peak FWHM, obtained from the
two crossings of the luminosity density, with linear
interpolation between bin centers. These widths are measured directly
from the spectra.

The points near 1 MeV use 80k macroparticles per beam and transverse
grids of $16384\times128$ cells for ILC/LCF and $16384\times256$ for
HALHF, with 96 and 48 longitudinal slices, respectively. The narrowest
points use 160k macroparticles per beam, $32768\times128$ and
$32768\times256$ cells, and 128 and 72 slices. Their analysis bin widths
are 0.025 MeV, compared with 0.05 MeV at the points near 1 MeV.
Each of these four settings has three independent collision seeds.
The reported FWHM is the mean of the per-seed widths and uncertainties are
standard errors of that mean. Sensitivity to bin alignment and doubling
the analysis bin width is reported so that binning effects are transparent.

\begin{table}[htbp]
\centering\small
\caption{Fixed optics horizontal dispersion scan at $W_0=125\GeV$. Luminosity is in units of $10^{32}\,\mathrm{cm}^{-2}\mathrm{s}^{-1}$ per interaction point. The final column is the
largest absolute fractional FWHM change over the tested bin alignments
and two analysis bin widths, taking the maximum over individual seeds. 
ILC/LCF luminosity uses the ILC 5 Hz repetition rate. The quoted uncertainties on the FWHM values represent statistical uncertainties across simulation seeds only.}
\label{tab:highd}
\begin{tabular}{@{}lrrrrr@{}}
\toprule
Family & $|D|$ [mm] & Seeds & $\mathcal L/10^{32}$ & FWHM [MeV] & Binning [\%]\\
\midrule
ILC/LCF & 0.000 & 3 & $93.2\pm5.9$ & $424.4\pm3.5$ & 0.47\\
ILC/LCF & 8.051 & 1 & 1.653 & 9.630 & 3.14\\
ILC/LCF & 16.107 & 1 & 0.726 & 4.871 & 1.93\\
ILC/LCF & 32.216 & 3 & $0.34130\pm0.00043$ & $2.5118\pm0.0077$ & 1.80\\
ILC/LCF & 75.865 & 3 & $0.13850\pm0.00035$ & $1.0695\pm0.0042$ & 1.88\\
ILC/LCF & 151.731 & 3 & $0.068441\pm0.000052$ & $0.5964\pm0.0017$ & 1.40\\
HALHF & 0.000 & 3 & $144.72\pm0.53$ & $350.3\pm1.4$ & 0.77\\
HALHF & 9.936 & 1 & 1.981 & 9.421 & 4.31\\
HALHF & 19.878 & 1 & 0.951 & 4.671 & 3.10\\
HALHF & 39.759 & 3 & $0.4694\pm0.0020$ & $2.433\pm0.027$ & 3.18\\
HALHF & 93.629 & 3 & $0.19395\pm0.00052$ & $1.0276\pm0.0058$ & 1.57\\
HALHF & 187.258 & 3 & $0.093141\pm0.000070$ & $0.5405\pm0.0060$ & 2.28\\
\bottomrule
\end{tabular}
\end{table}

The narrowest ILC/LCF and HALHF peaks have FWHMs of
$(0.5964\pm0.0017)\MeV$ at $|D|=151.731$ mm and
$(0.5405\pm0.0060)\MeV$ at $|D|=187.258$ mm, respectively.
Their instantaneous luminosities are
$(6.8441\pm0.0052)\times10^{30}$ and
$(9.3141\pm0.0070)\times10^{30}\,\mathrm{cm}^{-2}\mathrm{s}^{-1}$
respectively and per interaction point. Ratios of the mean luminosities to the corresponding
zero dispersion controls are approximately $0.073\%$ and $0.064\%$. Both linear collider machines achieve widths capable of Higgs resonance measurement at roughly 0.1\% of their nominal instantaneous luminosity. Figures~\ref{fig:ilcmono} and~\ref{fig:halhfmono} include these narrow peaks alongside the reference and luminosity conserving configurations.
Each displayed spectrum is normalized to its own maximum bin height, while Fig.~\ref{fig:dispersion} retains the absolute luminosities.

Peaks with widths smaller than 1~MeV are obtained in these collision configurations but at a substantial cost to instantaneous luminosity. The width alone does not establish the
sensitivity of a Higgs measurement, which also depends on statistics
near the resonance, initial state radiation, and beam-energy calibration
\cite{higgsresonance2022,higgslineshape2016}. Follow-up studies on these other factors are needed.

\subsection{FCC-ee with finite angle crab-waist scheme}
Here we investigate monochromatization with an FCC-ee where it retains a 30 mrad full crossing angle and a 90\% local crab-waist map. 
The latter shifts the vertical waist along the opposing beam trajectory and includes the conjugate horizontal kick. It represents the local canonical transformation rather than transport through a complete sextupole system.

The incoming parameters are the published MonochroM ZH2IP equilibrium
moments including beamstrahlung~\cite{fccmono2025}:
$N=5.54\times10^{10}$ particles per bunch,
$(\epsilon_x,\epsilon_y)=(3.56\,\mathrm{nm},3.42\,\mathrm{pm})$,
$(\beta_x^*,\beta_y^*)=(90,1)$ mm, $s=6.4\times10^{-4}$ and
$\sigma_z=4.74$ mm. Here the emittances are geometric and $s$ is the
fractional RMS energy spread. The collision rate per IP is
$3.9533\times10^7\,\mathrm{s}^{-1}$, corresponding to 12k bunches
in a 91 km ring. The beam energy is
$E_0=62.5/\cos(15\,\mathrm{mrad})=62.5070\GeV$, giving
$W_0=125\GeV$. We hold these moments fixed while performing an opposite-sign horizontal dispersion scanning up to $|D|=175$ mm.

\begin{table}[htbp]
\centering\small
\caption{FCC-ee finite angle crab-waist scan at
$W_0=125\GeV$. Luminosity is per IP. Below 100 mm, each point
uses one seed, 10k macroparticles per beam, $512\times4096$
transverse cells and 32 slices. From 100 mm, each point uses three seeds,
80k macroparticles per beam, $2048\times4096$ cells, 128 slices
and larger spatial domains.}
\label{tab:fcc-luminosity}
\begin{tabular}{@{}rrrr@{}}
\toprule
$|D|$ [mm] & $\mathcal L$ [$10^{34}$ cm$^{-2}$ s$^{-1}$] & FWHM [MeV] & Binning range [MeV]\\
\midrule
0 & 20.099 & 130.93 & [130.00, 132.93]\\
25 & 18.453 & 128.56 & [127.22, 130.90]\\
50 & 15.052 & 109.84 & [108.32, 111.93]\\
75 & 12.004 & 94.28 & [93.41, 97.08]\\
100 & $9.803\pm0.018$ & $81.14\pm0.24$ & [81.03, 82.89]\\
125 & $7.986\pm0.021$ & $71.22\pm0.15$ & [71.09, 73.46]\\
150 & $6.625\pm0.017$ & $63.29\pm0.17$ & [63.07, 65.49]\\
175 & $5.566\pm0.015$ & $56.985\pm0.028$ & [56.88, 59.42]\\
\bottomrule
\end{tabular}
\end{table}

\begin{figure}[p]
\centering\includegraphics[width=\textwidth]{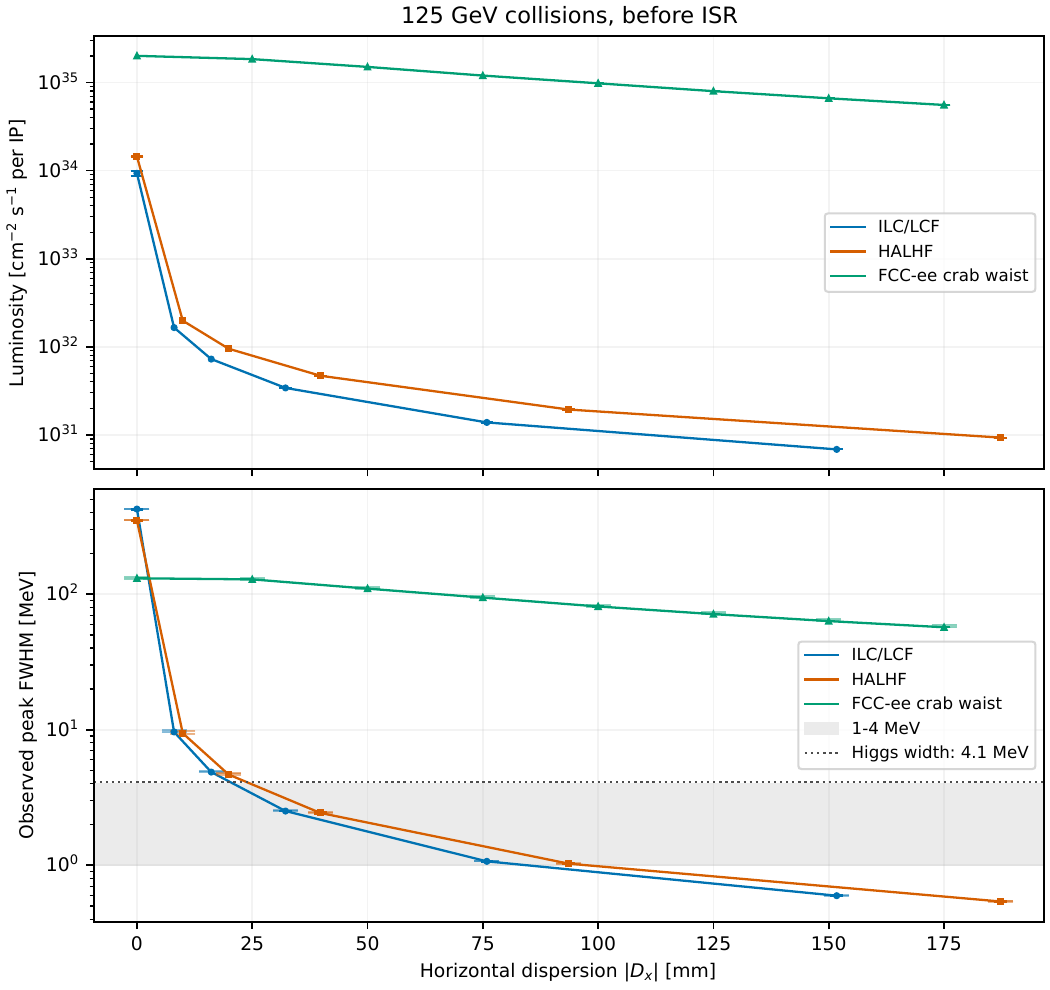}
\caption{Instantaneous luminosity per IP and observed peak FWHM versus
opposite horizontal dispersion at $W_0=125\GeV$.
Lines connect the simulated points. Seeds and run information are given in Tables~\ref{tab:highd} and~\ref{tab:fcc-luminosity}. The shaded interval is 1-4 MeV and the dotted line marks $\Gamma_H\approx4.1\MeV$. ILC/LCF luminosity uses the ILC 5 Hz rate.}
\label{fig:dispersion}
\end{figure}

Table~\ref{tab:fcc-luminosity} and Fig.~\ref{fig:dispersion} show a smooth
decrease in both luminosity and width. Between 100 and 175 mm, the FWHM
decreases from $(81.14\pm0.24)\MeV$ to $(56.985\pm0.028)\MeV$,
while luminosity decreases from $(9.803\pm0.018)\times10^{34}$ to
$(5.566\pm0.015)\times10^{34}\,\mathrm{cm}^{-2}\mathrm{s}^{-1}$.
Using the same three seeds across these settings gives paired decreases
of $(29.77\pm0.17)\%$ in width and $(43.224\pm0.070)\%$ in
luminosity. The 175 mm peak remains about fourteen times wider than
4 MeV. Therefore, even larger values of horizontal dispersion are needed for FCC-ee in such a configuration to reach the requirements for a direct Higgs measurement.

\subsection{Rigid Gaussian overlap comparison and projection}
\begin{figure}[p]
\centering\includegraphics[width=\textwidth]{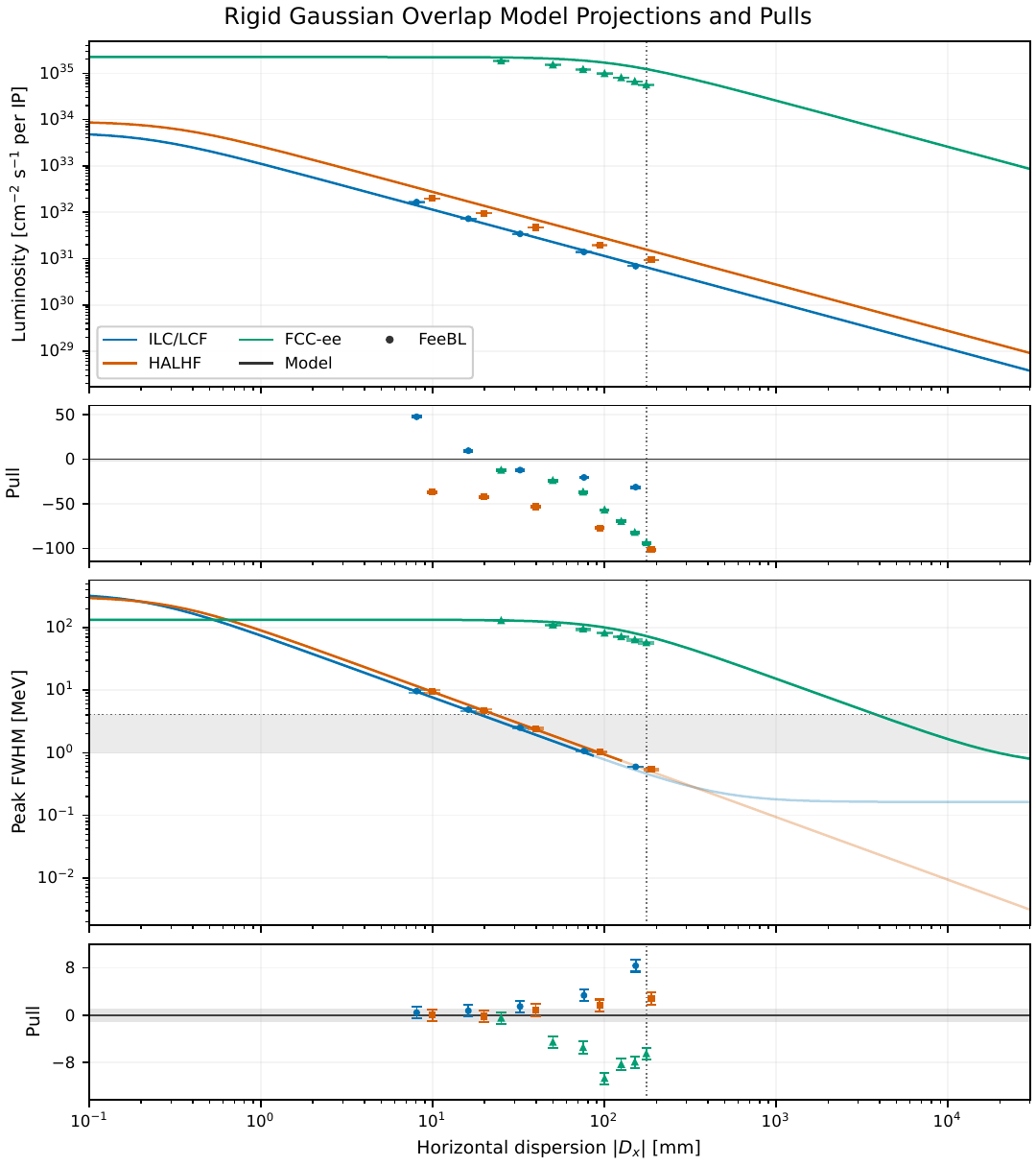}
\caption{Rigid Gaussian-overlap projections (lines) and \FeeBL\ simulations
(filled markers) at matching configurations. Luminosity and FWHM each
have a pull panel. The vertical line marks the FCC-ee simulation endpoint of 175 mm. The width band is 1-4 MeV and the horizontal dotted line is $\Gamma_H\approx4.1\MeV$.}
\label{fig:forecasts}
\end{figure}

Figure~\ref{fig:forecasts} compares the simulated dispersion scans with
analytical projections at matching incoming parameters. The model uses
$\beta_x^*=3.25$ mm for ILC/LCF, $0.825$ mm for HALHF and 90 mm for
FCC-ee, together with the energies, populations, emittances, energy
spreads, and collision rates of the simulated scans. ILC retains a
14 mrad crossing angle with ideal spatial tilt, HALHF is head-on, and
FCC-ee has a 30 mrad crossing angle without RF tilt. No parameters are
fitted to the simulation, and thus the pulls are not robust fitted model predictions, but instead, they are demonstrative of a purely analytical model. Here we use the term ``rigid'' to mean that the assigned Gaussian bunch densities do not evolve during the crossing.

For equal beams with opposite dispersions $D_1=-D_2=D$ and no spatial
tilt, the finite angle Gaussian overlap result
\cite{bogomyagkov2017,monoprinciple} is
\begin{align}
 A_x&=\epsilon_x\beta_x^*+\sigma_z^2\tan^2(\vca),
 &\Lambda_c^2&=1+\frac{D^2s^2}{A_x},\nonumber\\
 \mathcal L_{\rm rigid}
 &=\frac{f_{\rm BX}N^2}
 {4\pi\sqrt{A_x+D^2s^2}\sqrt{\epsilon_y\beta_y^*}},
 &\sigma_{W,\rm lin}^2
 &=\frac{W_0^2}{2}\left[
 \frac{s^2}{\Lambda_c^2}
 +\frac{\epsilon_x}{\beta_x^*}\tan^2(\vca)\right].
 \label{eq:rigid}
\end{align}

The quantity $A_x$ has units of length squared. SI lengths
give luminosity in $\mathrm{m}^{-2}\mathrm{s}^{-1}$; division by
$10^4$ gives $\mathrm{cm}^{-2}\mathrm{s}^{-1}$.
The unequal beam covariance calculation supplies the ILC/LCF and HALHF
curves, including the unequal ILC input energy spreads. The plotted model width is the first order Gaussian FWHM,$2\sqrt{2\ln2}\,\sigma_{W,\rm lin}$ while the simulated width is the
observed histogram FWHM.

The model reproduces the narrowing trend but does not reproduce all
absolute rates and widths within the estimated simulation uncertainties.
At the largest simulated dispersions, the ILC/LCF, HALHF and FCC-ee
luminosities are respectively 8.4\%, 36.3\% and 54.9\% below their
matched analytical predictions. The corresponding observed widths are
13.4\% above, 8.1\% above and 21.9\% below the model predictions.
Their width pulls are approximately $+8.4$, $+2.8$ and $-6.5$.
These differences compare a rigid Gaussian model with evolving,
radiating beams. Hourglass and crab-waist weighting, canonical transport
correlations and disruption are also absent from the analytical model.

The projections do not establish a realizable final focus or storage-ring
equilibrium. Ideal spatial tilt also omits RF amplitude, timing and
transport errors demonstrated in other studies~\cite{ilccrab2023}. The present study considers only opposite horizontal dispersion. Vertical dispersion, longitudinal
energy-position correlations, chromatic waist matching and combined
schemes~\cite{bogomyagkov2017,shatilov2018mono,fccmono2025,chromaticwaist2024}
are reserved for subsequent studies.
\FloatBarrier

\section{Program and data release}
The GitHub release accompanying this is listed as v0.7.0 in the GitHub releases and contains the source code, runnable cards, example inputs and outputs, digitized reference tables, and analysis scripts~\cite{github}. Data for full reproduction can be provided upon request.
\FloatBarrier
\section{Conclusions}

We have presented \FeeBL\ , a beam-beam simulation framework for electron-positron colliders, and investigated several effects relevant to precision experimental measurements. Comparisons with published LEP, ILC, HALHF and FCC-ee calculations demonstrate agreement with a range of existing predictions, while also identifying discrepancies and limitations that motivate further investigation and development work. The framework provides a means of studying luminosity spectra, beam electromagnetic deflection, beam pair production and beam (de)polarization.

For LEP, we reproduce the analytical field-integration calculations of small angle Bhabha scattering and find that previously published \GP simulations underestimate the corresponding angular deflections by approximately 2\%. Applying a revised beam-deflection correction, together with a data driven reassessment of the uncertainties among the four LEP experiments, gives $N_\nu=2.9965\pm0.0040$. The revised uncertainty requires a future reassessment of the experiment specific and correlated uncertainty budgets and the theory modeling. This result motivates future studies into improvements of the original studies to realize a potentially improved measurement.

We have identified and corrected several implementation errors in the inherited \GP++ beam pair calculations, including the treatment of forward-angle sampling, process selection, normalization and beam pair counting. Comparisons of Breit-Wheeler, Bethe-Heitler and Landau-Lifshitz processes demonstrate the sensitivity of predicted backgrounds to the approximations and prescriptions employed. These results are relevant to detector-background studies at future colliders, although further comparisons and studies remain necessary. Spin-tracking calculations reproduce the approximate scale of published ILC depolarization predictions and demonstrate a reduction in luminosity weighted depolarization for collisions near the nominal center-of-mass energy.

Finally, we have investigated monochromatization via horizontal dispersion at a collision energy of 125 GeV, exploring the direct Higgs measurement. For the exploratory ILC/LCF and HALHF configurations, the simulations demonstrate a $\approx2\times$ reduction in the fitted collision energy response width while conserving total luminosity and enhancing the nominal luminosity within the peak region. Larger dispersions produce sub-MeV peak widths, although the corresponding instantaneous luminosities fall to approximately 0.1\% of the zero dispersion configurations. In contrast, the FCC-ee configurations studied exhibit limited narrowing, and the analytical projections indicate substantial challenges for achieving widths comparable to the Higgs width using horizontal dispersion alone at FCC-ee.

The results presented here illustrate the importance of accurate beam-beam modeling for precision measurements. Further development of \FeeBL\ will focus on improved convergence, multiple GPU support, additional beam pair processes and higher order corrections, additional monochromatization schemes amongst other things. The release of the simulation software and benchmark data is intended to facilitate independent validation and further studies of beam induced effects at existing and future electron-positron colliders.

\FloatBarrier
\section{Acknowledgments}
We acknowledge that agentic AI was used in the writing and reviewing of programs, examples, and supplementary program documentation for this work, though we warrant that everything was checked and verified by a human researcher afterwards and remade when needed. All written text in this work is written by a human researcher though some content was started by agentic AI and replaced with human written text later. Some of the LaTex source code used to render this document were written with agentic AI. Mathematical derivations were done both by multiple forms of agentic AI and a human researcher to ensure correctness. We thank Wolfram Research for providing a complimentary Mathematica license that was used in the research published here. We thank Daniel Schulte for discussions and existing work on \GP. We thank Jenny List, Mikael Berggren and Shriyansh Ranjan for discussions and suggestions related to this work. We thank Ivo Schulthess for a review of a draft of this paper. The author acknowledges support from Deutsches Elektronen-Synchrotron (DESY), a member of the Helmholtz
Association (HGF). This research utilized
computational resources provided by the Maxwell
high-performance computing cluster and the German
National Analysis Facility (NAF), both operated
at DESY, Hamburg, Germany.

\newpage

\end{document}